\documentclass[12pt,preprint]{aastex}
\usepackage{amsmath,amsfonts,amssymb}

\begin{document}
\title{Generalized Chaplygin Gas Models tested with SNIa}
\vspace{0.6 cm}
\author{ {\sc Marek Biesiada} \\
{\sl Department of Astrophysics and  Cosmology, } \\
{\sl University of Silesia}\\
{\sl Uniwersytecka 4,  40-007 Katowice, Poland} \\
{mb@imp.sosnowiec.pl}\\
{\sc W{\l}odzimierz God{\l}owski} \\
{\sl Astronomical Observatory} \\
{\sl Jagiellonian University }\\
{\sl Orla171,  Krakow, Poland}  \\
{godlows@oa.uj.edu.pl} \\
{\sc Marek Szyd{\l}owski} \\
{\sl Astronomical Observatory} \\
{\sl Jagiellonian University }\\
{\sl Orla171,  Krakow, Poland}  \\
{szydlo@oa.uj.edu.pl}\\ }
\date{}
 
\begin{abstract}
\noindent The  Generalized Chaplygin Gas (GCG) with the equation of state 
$p = - \frac{A}{{\rho}^{\alpha}}$ was recently proposed as a candidate for 
dark energy in the Universe. In this paper we confront the GCG with SNIa data 
using avaliable samples. Specifically we have tested the GCG
cosmology in three different classes of models with (1) $\Omega_m= 0.3$,
$\Omega_{Ch}= 0.7$; (2) $\Omega_m= 0.05$, $\Omega_{Ch}= 0.95$ and
(3) $\Omega_m = 0$, $\Omega_{Ch} =  1$, as well as a model without prior
assumptions on $\Omega_m$. The best fitted models are obtained by
minimalizing the $\chi^2$ function. We supplement our analysis with confidence
intervals in the $(A_0, \alpha)$ plane by marginalizing the probability
density functions over remaining parameters assuming uniform priors.
We have also derived one-dimensional probability distribution functions for
$\Omega_{Ch}$ obtained from joint marginalization over $\alpha$ and $A_0$.
The maximum value of such PDF provides the most probable value of 
$\Omega_{Ch}$ within the full class of GCG models.
The general conclusion is that SNIa data give support to the Chaplygin gas
(with $\alpha =  1$). However noticeable preference of $A_0$ values close to
1 means that the  $\alpha$ dependence becomes insignificant. It is reflected
on one dimensional PDFs for $\alpha$ which turned out to be flat meaning that
the power of present supernovae data to discriminate between various GCG
models (differing by $\alpha$) is weak. Extending our analysis by relaxing
the prior assumption of the flatness of the Universe leads to the result
that even though the best fitted values of $\Omega_k$ are formally non-zero,
still they are close to the flat case. 
Our results show clearly that in GCG cosmology distant (i.e. $z >1$)
supernovae should be brighter than in $\Lambda$CDM model. Therefore one can 
expect that future supernova experiments (e.g., SNAP) having access to higher 
redshifts will eventually resolve the issue whether the dark energy content 
of the Universe could be described as a the Chaplygin gas.
Moreover, it would be possible to differentiate between models with various 
value of $\alpha$ parameter and/or discriminated between GCG, Cardassian and 
$\Lambda$CDM models. This discriminative power of the forthcoming mission has 
been demonstrated on simulated SNAP data.

\end{abstract}
 
\maketitle
 
\section{Introduction}
 
For a couple of years two independent observational programs --- the high
redshift supernovae surveys (Perlmutter et al. 1999) and CMBR small scale
anisotropy measurements (de Bernardis et al. 2000, Benoit et al. 2003,
Hinshaw et al. 2003) have brought a new picture of the Universe in the large.
While interpreted  ithin the FRW models results of these programs suggest
that our Universe is flat (as inferred from the location of acoustic peaks
in CMBR power spectrum) and presently accelerates its expansion (as inferred
from the SNIa Hubble diagram). Combined with the independent knowledge about
the amount of baryons and CDM estimated to be $\Omega_m =  0.3$ (Turner 2002)
it follows that about $\Omega_X =  0.7$ fraction of critical density
$\rho_{cr} =  \frac{3 c^2 H_0^2}{8 \pi G}$ should be contained in a mysterious
component called ``dark energy''. The most obvious candidate for this smooth
component permeating the Universe is the cosmological constant $\Lambda$
representing the energy of the vacuum. Well known fine tuning problems
led many people to seek beyond the $\Lambda$ framework, and the concept of
the quintessence had been conceived. Usually the quintessence is described
in a phenomenological manner, as a scalar  field with an appropriate
potential (Ratra \& Peebles 1988, Caldwell, Dave \& Steinhardt 1995, Frieman,
Stebbins \& Waga 1995). It turns out, however, that quintessence program also
suffers from its own fine tuning problems (Kolda \& Lyth 1999).
 
In 1904 Russian physicist Chaplygin introduced the exotic equation of
state $p =  - \frac{A}{\rho}$ to discribe an adiabatic aerodynamic process
(Chaplygin 2004). The attractiveness of this equation of state in the
context dark energy models comes mainly from the fact that it gives the
unification of both dark energy (postulated in cosmology to explain current 
aceleration of the Universe) 
 and clustered dark matter which is postulated in
astrophysics to explain the flat rotation curves of spiral galaxies.
It is interesting that the Chaplygin gas can be derived from the quintessence
Lagrangian for the scalar field $\phi$ with some potential and also from the
Born-Infeld form of the Lagrangian (Kamenshchik, Moschella \& Pasquier 2001).
The Chaplygin equation of state has some interesting connections with string
theory and it admits the interpretation in the framework of 
brane cosmologies (Jackiw 2000).
 
Recently the so called Chaplygin gas (Kamenshchik, Moschella \& Pasquier 2001,
Fabris, Gon{\c c}alves \& de Souza 2002, Szyd{\l}owski \& Czaja 2004) --- a
hypothetical component with the equation of state $p =  - \frac{A}{\rho}$ ---
was proposed as a challenge to the above mentioned candidates for dark energy.
This, also purely phenomenological, entity has interesting connections with
string theory (Ogawa 2000). Currently its generalizations admitting the
equation of state $p=  - \frac{A}{\rho^{\alpha}}$ where $0\le \alpha \le 1$
have been proposed (Bento, Bertolami \& Sen 2002, Carturan \& Finelli 2002a).
 
In this paper we confront the Generalized Chaplygin Gas with the SNIa data.
At this point our choice of Generalized Chaplygin Gas cosmologies deserves
a sort of justification. There are two approaches in the literature. First
one is phenomenological, namely having no preferred theory of dark energy
responsible for acceleration of the Universe one characterizes it as a
cosmic fluid  with an equation of state $p_X = w \rho_X$ where
$w \geq -1$ (see e.g. (Chiba 1998,Turner \& White 1997) and an immense
literature that appeared thereafter). Because, as already mentioned above,
a strain of ideas about dark energy is associated with an evolving scalar
there are good reasons to expect that cosmic equation of state could be
time dependent i.e. $ w = w(t) = w(z)$ (e.g. Weller \& Albrecht 2001,
Maor et al. 2001 and many others thereafter). This approach seems attractive
from the perspective of analyzing observational data such like supernovae
surveys and indeed this approach was taken while first analyzing the data
(Perlmutter et al. 1999, Knop et al. 2003 or Riess et al. 2004). However
even though such analysis places constrains on {\it any} potential theory
that might explain the dark energy phenomenon, ultimately one always ends up
at testing a {\it specific} theory. Along this line there appeared attempts
to reconstruct the scalar potential, if the scalar field was responsible for
dark energy (e.g. Alam et al. 2003 and references therein).
Our approach goes along this philosophy but instead is devoted to the
Generalized Chaplygin Gas which is being recently considered as candidate
to unified dark matter-energy component (i.e. responsible for both clustering
and accelerated expansion (Makler, de Oliveira \& Waga 2003).
 
The cosmological models with the Generalized Chaplygin Gas have also many
special features which make them atractive. In standard cosmological model
one can clearly distinguish the epochs of radiation domination followed by
(ordinary) matter domination (with decelerated expansion). As mentioned above
supernova data suggest that the epoch of decelerated expansion ended and
switched to accelerated epoch --- dominated by dark energy. The Generalized
Chaplygin Gas models describe smoothly the transition from the decelerated
to accelerated epochs. They represent the simplest deformation of concordance
$\Lambda$CDM (Gorini et al. 2004). And moreover, they propose a new unified
macroscopic (phenomenological) description of both dark energy and dark matter.
This places them in a distinguished position from the point of view of Occam's
razor principle. It should be also noted that the Genaralized Chaplygin Gas
model allowed us to explain presently observed acceleration of the Universe
without the cosmological constant and/or modification of Einstein's equations.
 
If one takes seriously given dark energy scenario (necessary to explain
cosmic acceleration) one should also consider the behaviour of perturbations
in such a universe. In the framework of quintessence models with the barotropic
equation of state (i.e. $p= w \rho$ and $w=const$) one faces the problem of
instabilities in short scales. This appears because the speed of sound squared
(equal here to $w$) is negative (and constant). Calculation of the sound
speed in Generalized Chaplygin Gas model (see below) reveal its non-barotropic
nature. The perturbations in GCG models are stable in short scales even in
an accelerating phase (Carturan \& Finelli 2002a). Moreover, they behave
like dust perturbations when Chaplygin Gas is in dust regime.
 
Another motivation for studying Generalized Chaplygin Gas models goes from
theoretical physics --- specifically from attempts to describe the dark energy
in terms of the Lagrangian for a tachyonic field (Garousi 2000, Sen 2002).
Of course it would be nice to have a description of dark energy in terms
of the non quintessence Lagrangian as it  describes the nature of dark energy
while the cosmological constant is only phenomelogical and effective
description. One should also note that the Generalized Chaplygin Gas
equation of state arises in modern physics in the context of brane models
(Bordemann \& Hoppe 1993, Kamenshchik, Moschella \& Pasquier 2001,
Randall \& Sundrum 1999) where the Generalized Chaplygin Gas manifests
itself as an effect of immersion of our Universe in multidimensional
bulk space.
 
Generalized Chaplygin gas models have been intensively studied in the
literature and in particular they have been tested against supernovae data
(Makler, de Oliveira \& Waga 2003, Avelino et al. 2003,
Collistete et al. 2003), lensing statistics (Dev, Alcaniz \& Jain 2003),
CMBR measurements (Bento, Bertolami \& Sen 2003a, 20003b,
Carturan \& Finelli 2003b, Amendola et al. 2003), age-redshift relation
(Alcaniz, Jain \& Dev 2003), x-ray luminosities of galaxy clusters
(Cunha, Lima \& Alcaniz 2003) or from the large scale structure
considerations (Bean \& Dor{\'e} 2003, Multamaki, Manera \& Gaztanaga 2003,
Bilic et al. 2003). Perspectives to distinguish between Generalized
Chaplygin Gas, brane-world scenarios and quintessence in forthcoming
gravity wave experiments has been discussed in (Biesiada 2003). Although
the results are in general mutually consistent there was no strong
convergence to unique values of $A_0$, $\alpha$ parameters characterizing
Chaplygin gas equation of state.
 
Makler, de Oliveira \& Waga (2003) have considered the FRW model filled
completely with Generalized Chaplygin Gas and concluded that whole class
of such models is consistent with current SNIa data although the value of
$\alpha =  0.4$ is favoured. This result has been confirmed by our analysis
(class (3) models). However, when the existing knowledge about baryonic
matter content of the Universe was incorporated into the study our results
were different from Makler, de Oliveira \& Waga (2003) who found that
$\alpha =  0.15$ was preferred (assuming $\Omega_m =  0.04$ which is very
close to our assumption for class (2)  models).
 
As noticed by Bean \& Dor{\'e} (2003) Generalized Chaplygin Gas models have
an inherent degeneracy with cosmological constant models as far as background
evolution is concerned, and therefore they have a good fit with SNIa data.
These degeneracies disappear at the level of evolution of perturbations and
hence confrontation with CMBR spectrum would be decisive. Using available
data on the position of CMBR peaks measured by BOOMERANG
(de Bernardis et al. 2000) and Archeops (Benoit et al. 2003,
Hinshaw et al. 2003, Bento, Bertolami \& Sen (2002) obtained the following
constraints: $0.81 \leq A_0 \leq 0.85$ and $0.2 \leq \alpha \leq 0.6$ at
$68 \%$ CL in the model representative of our class (2) (i.e. with
$\Omega_m = 0.05$ assumed). Another estimation of the parameter $\alpha$
was done by Amendola et al. (2003) with WMAP Data. The obtained the
$0\le \alpha < 0.2$ at $95\%$ confidence level.

Using the angular size statistics for extragalactic sources combined with
SNIa data it was found in (Alcaniz \& Lima 2003) that in the the $\Omega_m =0.3$
and $\Omega_{Ch}=0.7$ scenario best fitted values of model parameters are
$A_0=0.83$ and $\alpha=1.$ respectively.
Recent paper by Bertolami et al. (2004) in which Generalized Chaplygin Gas
models have been analyzed against Tonry et al. (2003) supernovae data relaxing
the prior assumption on flatness suggests, surprisingly as the authors admit,
the preference of $\alpha > 1$.

\begin{deluxetable}{@{}p{1.5cm}rrrrrrr}
\tabletypesize{\scriptsize}
\tablewidth{0pt}
\tablecaption{Results of statistical analysis of Generalized Chaplygin Gas model
(with marginalization over ${\cal M}$) performed on analyzed
samples of SNIA (A, C, K6, K3, TBI, TBII, Silver, Gold) as a minimum $\chi^2$ best-fit
(denoted BF) and with the maximum likelihood method (denoted L). First two rows for
each sample refer to no prior on
$\Omega_m$. The same analysis was repeated with fixed priors
$\Omega_m=0.0$, $\Omega_m=0.05$ and $\Omega_m=0.3$.}
\label{tab:1}
\tablehead{\colhead{sample} & \colhead{$\Omega_m$} & \colhead{$\Omega_{Ch}$} &
\colhead{$A_0$} & \colhead{$\alpha$} & \colhead{$\mathcal{M}$} & \colhead{$\chi^2$}& \colhead{method}}
\startdata
A   &  0.00 & 1.00 & 0.77 & 1.00 &-3.39 & 95.4 &  BF  \\
    &  0.17 & 0.83 & 0.83 & 0.00 &-3.36 & ---  &  L   \\
    &  0.00 & 1.00 & 0.77 & 1.00 &-3.39 & 95.4 &  BF  \\
    &  0.00 & 1.00 & 0.73 & 1.00 &-3.38 & ---  &  L   \\
    &  0.05 & 0.95 & 0.80 & 1.00 &-3.39 & 95.4 &  BF  \\
    &  0.05 & 0.95 & 0.76 & 1.00 &-3.38 & ---  &  L   \\
    &  0.30 & 0.70 & 0.96 & 1.00 &-3.39 & 95.8 &  BF  \\
    &  0.30 & 0.70 & 0.96 & 0.00 &-3.38 & ---  &  L   \\
\tableline 
C   &  0.00 & 1.00 & 0.80 & 1.00 &-3.44 & 52.9 &  BF  \\
    &  0.15 & 0.85 & 0.86 & 0.00 &-3.41 & ---  &  L   \\
    &  0.00 & 1.00 & 0.80 & 1.00 &-3.44 & 52.9 &  BF  \\
    &  0.00 & 1.00 & 0.76 & 0.49 &-3.43 & ---  &  L   \\
    &  0.05 & 0.95 & 0.83 & 1.00 &-3.44 & 53.0 &  BF  \\
    &  0.05 & 0.95 & 0.79 & 0.11 &-3.43 & ---  &  L   \\
    &  0.30 & 0.70 & 0.99 & 1.00 &-3.42 & 53.3 &  BF  \\
    &  0.30 & 0.70 & 0.99 & 0.00 &-3.39 & ---  &  L   \\
\tableline
K6  &  0.00 & 1.00 & 0.81 & 1.00 &-3.52 & 55.3 &  BF  \\
    &  0.10 & 0.90 & 0.88 & 0.00 &-3.51 & ---  &  L   \\
    &  0.00 & 1.00 & 0.81 & 1.00 &-3.52 & 55.3 &  BF  \\
    &  0.00 & 1.00 & 0.78 & 0.71 &-3.52 & ---  &  L   \\
    &  0.05 & 0.95 & 0.84 & 1.00 &-3.52 & 55.4 &  BF  \\
    &  0.05 & 0.95 & 0.81 & 0.06 &-3.52 & ---  &  L   \\
    &  0.30 & 0.70 & 1.00 & 1.00 &-3.51 & 55.9 &  BF  \\
    &  0.30 & 0.70 & 1.00 & 0.00 &-3.49 & ---  &  L   \\
\tableline
K3  &  0.00 & 1.00 & 0.85 & 1.00 &-3.48 & 60.4 &  BF  \\
    &  0.11 & 0.89 & 0.88 & 0.00 &-3.45 & ---  &  L   \\
    &  0.00 & 1.00 & 0.85 & 1.00 &-3.48 & 60.4 &  BF  \\
    &  0.00 & 1.00 & 0.80 & 0.30 &-3.47 & ---  &  L   \\
    &  0.05 & 0.95 & 0.87 & 1.00 &-3.47 & 60.4 &  BF  \\
    &  0.05 & 0.95 & 0.84 & 0.00 &-3.47 & ---  &  L   \\
    &  0.30 & 0.70 & 1.00 & 1.00 &-3.44 & 61.4 &  BF  \\
    &  0.30 & 0.70 & 1.00 & 0.00 &-3.42 & ---  &  L   \\
\tableline
TBI &  0.00 & 1.00 & 0.79 & 1.00 &15.895&273.9 &  BF  \\
    &  0.00 & 1.00 & 0.81 & 1.00 &15.905& ---  &  L   \\
    &  0.00 & 1.00 & 0.79 & 1.00 &15.895&273.8 &  BF  \\
    &  0.00 & 1.00 & 0.75 & 1.00 &15.905& ---  &  L   \\
    &  0.05 & 0.95 & 0.82 & 1.00 &15.895&274.0 &  BF  \\
    &  0.05 & 0.95 & 0.78 & 1.00 &15.915& ---  &  L   \\
    &  0.30 & 0.70 & 0.97 & 1.00 &15.915&275.8 &  BF  \\
    &  0.30 & 0.70 & 0.96 & 0.00 &15.915& ---  &  L   \\
\tableline
TBII &  0.00 & 1.00 & 0.78 & 1.00 &15.915&186.5 &  BF  \\
     &  0.00 & 1.00 & 0.81 & 1.00 &15.925& ---  &  L   \\
     &  0.00 & 1.00 & 0.78 & 1.00 &15.915&186.5 &  BF  \\
     &  0.00 & 1.00 & 0.75 & 1.00 &15.915& ---  &  L   \\
     &  0.05 & 0.95 & 0.81 & 1.00 &15.915&186.6 &  BF  \\
     &  0.05 & 0.95 & 0.78 & 1.00 &15.925& ---  &  L   \\
     &  0.30 & 0.70 & 0.97 & 1.00 &15.925&188.4 &  BF  \\
     &  0.30 & 0.70 & 0.96 & 0.00 &15.935& ---  &  L   \\
\tableline
Silver&  0.00 & 1.00 & 0.82 & 1.00 &15.945&229.4 &  BF  \\
      &  0.00 & 1.00 & 0.84 & 1.00 &15.945& ---  &  L   \\
      &  0.00 & 1.00 & 0.82 & 1.00 &15.945&229.4 &  BF  \\
      &  0.00 & 1.00 & 0.79 & 1.00 &15.955& ---  &  L   \\
      &  0.05 & 0.95 & 0.85 & 1.00 &15.945&229.6 &  BF  \\
      &  0.05 & 0.95 & 0.81 & 1.00 &15.955& ---  &  L   \\
      &  0.30 & 0.70 & 0.99 & 1.00 &15.965&232.3 &  BF  \\
      &  0.30 & 0.70 & 0.99 & 0.00 &15.965& ---  &  L   \\
\tableline
Gold &  0.00 & 1.00 & 0.81 & 1.00 &15.945&173.7 &  BF  \\
       &  0.00 & 1.00 & 0.83 & 1.00 &15.955& ---  &  L   \\
       &  0.00 & 1.00 & 0.81 & 1.00 &15.945&173.7 &  BF  \\
       &  0.00 & 1.00 & 0.77 & 1.00 &15.955& ---  &  L   \\
       &  0.05 & 0.95 & 0.84 & 1.00 &15.945&173.8 &  BF  \\
       &  0.05 & 0.95 & 0.80 & 1.00 &15.955& ---  &  L   \\
       &  0.30 & 0.70 & 0.99 & 1.00 &15.965&175.6 &  BF  \\
       &  0.30 & 0.70 & 0.99 & 0.00 &15.965& ---  &  L   \\
\enddata
\end{deluxetable}

\begin{table}
\caption{Generalized Chaplygin Gas model parameter values obtained
from the marginal probability density functions calculated on Perlmutter,
Knop, Tonry/Barris and Riess samples with $\Omega_m$ prior relaxed.}
\label{tab:2}
\begin{tabular}{@{}p{1.5cm}rrrr}
\hline
\hline sample & $\Omega_m$ & $\Omega_{Ch}$ &  $A_0$ & $\alpha$ \\
\hline
A &  $ 0.17^{+0.08}_{-0.17}$ & $0.83^{+0.17}_{-0.08}$   & $0.83^{+0.14}_{-0.09}$ & $-0.0^{+0.67}$ \\
C & $0.15^{+0.08}_{-0.15}$ & $0.85^{+0.15}_{-0.08}$   & $0.86^{+0.13}_{-0.10}$ & $0.0^{+0.66}$  \\
K6&  $ 0.10^{+0.11}_{-0.10}$ & $0.90^{+0.10}_{-0.11}$   & $0.88^{+0.12}_{-0.08}$ & $-0.0^{+0.66}$ \\
K3& $0.11^{+0.07}_{-0.11}$ & $0.89^{+0.11}_{-0.07}$   & $0.88^{+0.11}_{-0.05}$ & $0.0^{+0.66}$\\
TBI& $0.00^{+0.21}$ & $1.00_{-0.21}$   & $0.81^{+0.12}_{-0.07}$ & $1.0_{-0.60}$\\
TBII&$0.00^{+0.21}$ & $1.00_{-0.21}$   & $0.81^{+0.12}_{-0.07}$ & $1.0_{-0.62}$\\
Silver&$0.00^{+0.18}$ & $1.00_{-0.18}$   & $0.84^{+0.09}_{-0.06}$ & $1.0_{-0.59}$\\
Gold&  $0.00^{+0.20}$ & $1.00_{-0.20}$   & $0.83^{+0.11}_{-0.07}$ & $1.0_{-0.64}$\\
\hline
\end{tabular}
\end{table}

\begin{table}
\caption{Generalized Chaplygin Gas model parameter values obtained from the
marginal probability density functions calculated on Perlmutter, Knop,
Tonry/Barris and Riess samples.  The analysis was done with
fixed $\Omega_m=0.0$, $\Omega_m=0.05$ and $\Omega_m=0.3$.}
\label{tab:3}
\begin{tabular}{@{}p{1.5cm}rrrr}
\hline
\hline sample & $\Omega_m$ & $\Omega_{Ch}$ &  $A_0$ & $\alpha$ \\
\hline
A     &$ 0.00$ & $1.00$ & $0.73^{+0.08}_{-0.10}$ & $1.0_{-0.63}$ \\
      &$ 0.05$ & $0.95$ & $0.76^{+0.08}_{-0.09}$ & $1.0_{-0.66}$ \\
      &$ 0.30$ & $0.70$ & $0.96^{+0.04}_{-0.09}$ & $0.0^{+0.65}$ \\
C     &$ 0.00$ & $1.00$ & $0.76^{+0.08}_{-0.10}$ & $0.49^{+0.36}_{-0.35}$ \\
      &$ 0.05$ & $0.95$ & $0.79^{+0.08}_{-0.11}$ & $0.41^{+0.27}_{-0.41}$ \\
      &$ 0.30$ & $0.70$ & $0.99^{+0.01}_{-0.11}$ & $0.0^{+0.64}$ \\
K6    &$ 0.00$ & $1.00$ & $0.78^{+0.07}_{-0.09}$ & $0.71^{+0.29}_{-0.40}$ \\
      &$ 0.05$ & $0.95$ & $0.81^{+0.08}_{-0.09}$ & $0.06^{+0.61}_{-0.06}$ \\
      &$ 0.30$ & $0.70$ & $1.00_{-0.10}        $ & $0.0^{+0.64}$ \\
K3    &$ 0.00$ & $1.00$ & $0.80^{+0.06}_{-0.06}$ & $0.30^{+0.39}_{-0.30}$ \\
      &$ 0.05$ & $0.95$ & $0.84^{+0.05}_{-0.07}$ & $0.0^{+0.67}$ \\
      &$ 0.30$ & $0.70$ & $1.00_{-0.06}        $ & $0.0^{+0.63}$ \\
TBI   &$ 0.00$ & $1.00$ & $0.75^{+0.04}_{-0.05}$ & $1.0_{-0.54}$ \\
      &$ 0.05$ & $0.95$ & $0.78^{+0.04}_{-0.06}$ & $1.0_{-0.55}$ \\
      &$ 0.30$ & $0.70$ & $0.96^{+0.04}_{-0.04}$ & $0.0^{+0.67}$ \\
TBII  &$ 0.00$ & $1.00$ & $0.75^{+0.04}_{-0.06}$ & $1.0_{-0.54}$ \\
      &$ 0.05$ & $0.95$ & $0.78^{+0.04}_{-0.06}$ & $1.0_{-0.54}$ \\
      &$ 0.30$ & $0.70$ & $0.96^{+0.04}_{-0.04}$ & $0.0^{+0.67}$ \\
Silver&$ 0.00$ & $1.00$ & $0.79^{+0.03}_{-0.05}$ & $1.0_{-0.52}$ \\
      &$ 0.05$ & $0.95$ & $0.81^{+0.04}_{-0.04}$ & $1.0_{-0.54}$ \\
      &$ 0.30$ & $0.70$ & $0.99^{+0.01}_{-0.03}$ & $0.0^{+0.64}$ \\
Gold  &$ 0.00$ & $1.00$ & $0.77^{+0.04}_{-0.05}$ & $1.0_{-0.58}$ \\
      &$ 0.05$ & $0.95$ & $0.80^{+0.04}_{-0.05}$ & $1.0_{-0.59}$ \\
      &$ 0.30$ & $0.70$ & $0.99^{+0.01}_{-0.04}$ & $0.0^{+0.64}$ \\
\hline
\end{tabular}
\end{table}
 
\begin{table}
\noindent
\caption{Results of statistical analysis of Generalized Chaplygin Gas
models with flat prior relaxed and with marginalization over ${\cal M}$
performed on Knop Samples K3 as a minimum $\chi^2$ best-fit(denoted BF)
and with the maximum likelihood method (denoted L). First two rows refer to
no prior on $\Omega_m$. The same analysis was repeated with
fixed $\Omega_m=0.0$, $\Omega_m=0.05$ and $\Omega_m=0.3$.}
\label{tab:4}
\begin{tabular}{@{}p{1.5cm}rrrrrrrr}
\hline
\hline sample & $\Omega_k$ & $\Omega_m$ & $\Omega_{Ch}$ &  $A_0$ & $\alpha$ & $\mathcal{M}$ & $\chi^2$& method \\
\hline
K3  & -0.19  &  0.00 & 1.19 & 0.82 & 1.00 &-3.48 & 60.3 &  BF  \\
    & -0.60  &  0.00 & 1.26 & 0.89 & 0.00 &-3.46 & ---  &  L   \\
    & -0.25  &  0.00 & 1.25 & 0.82 & 1.00 &-3.49 & 60.3 &  BF  \\
    &  0.10  &  0.00 & 0.90 & 0.76 & 0.00 &-3.46 & ---  &  L   \\
    & -0.28  &  0.05 & 1.23 & 0.84 & 1.00 &-3.49 & 60.3 &  BF  \\
    &  0.05  &  0.05 & 0.90 & 0.78 & 0.00 &-3.47 & ---  &  L   \\
    & -0.48  &  0.30 & 1.18 & 0.93 & 0.97 &-3.49 & 60.3 &  BF  \\
    & -0.35  &  0.30 & 1.05 & 0.88 & 0.00 &-3.47 & ---  &  L   \\
Gold& -0.12  &  0.00 & 1.12 & 0.80 & 0.99 &15.945&173.4 &  BF  \\
    & -0.32  &  0.00 & 1.06 & 0.82 & 0.00 &15.945& ---  &  L   \\
    & -0.13  &  0.00 & 1.13 & 0.81 & 1.00 &15.935&173.4 &  BF  \\
    & -0.19  &  0.00 & 1.19 & 0.76 & 0.85 &15.945& ---  &  L   \\
    & -0.17  &  0.05 & 1.12 & 0.83 & 1.00 &15.935&173.4 &  BF  \\
    & -0.20  &  0.05 & 1.15 & 0.78 & 0.54 &15.945& ---  &  L   \\
    & -0.31  &  0.30 & 1.01 & 0.94 & 1.00 &15.955&173.6 &  BF  \\
    & -0.30  &  0.30 & 1.00 & 0.91 & 0.00 &15.945& ---  &  L   \\
\hline
\end{tabular}
\end{table}

\begin{table}
\caption{Results of statistical analysis of Generalized Chaplygin Gas models
with flat prior relaxed and with
marginalization over ${\cal M}$ performed on Knop Samples K3.
Model parameter values are obtained
from the marginal probability density functions.
First row refer to
no prior on $\Omega_m$. The same analysis was repeated with
fixed $\Omega_m=0.0$, $\Omega_m=0.05$ and $\Omega_m=0.3$.}
\label{tab:5}
\begin{tabular}{@{}p{1.5cm}rrrrr}
\hline
\hline sample & $\Omega_k$ & $\Omega_m$ & $\Omega_{Ch}$ &  $A_0$ & $\alpha$ \\
\hline
K3  & $-0.60^{+0.38}_{-0.38}$ & $ 0.00^{+0.29}$ & $1.26^{+0.25}_{-0.39}$ & $0.89^{+0.11}_{-0.07}$ & $ 0.0^{+0.64}$ \\
    & $ 0.10^{+0.37}_{-0.60}$ & $ 0.00$         & $0.90^{+0.59}_{-0.37}$ & $0.76^{+0.10}_{-0.07}$ & $ 0.0^{+0.66}$  \\
    & $ 0.05^{+0.31}_{-0.58}$ & $ 0.05$         & $0.90^{+0.58}_{-0.31}$ & $0.78^{+0.10}_{-0.06}$ & $ 0.0^{+0.66}$ \\
    & $-0.35^{+0.17}_{-0.40}$ & $ 0.30$         & $1.05^{+0.41}_{-0.17}$ & $0.88^{+0.09}_{-0.05}$ & $ 0.0^{+0.63}$  \\
Gold& $-0.32^{+0.25}_{-0.25}$ & $ 0.00^{+0.28}$ & $1.06^{+0.24}_{-0.22}$ & $0.82^{+0.13}_{-0.05}$ & $ 0.0^{+0.64}$ \\
    & $-0.19^{+0.29}_{-0.28}$ & $ 0.00$         & $1.19^{+0.28}_{-0.29}$ & $0.76^{+0.03}_{-0.05}$ & $ 0.85^{+0.15}_{-0.52}$  \\
    & $-0.20^{+0.28}_{-0.29}$ & $ 0.05$         & $1.15^{+0.29}_{-0.28}$ & $0.78^{+0.05}_{-0.06}$ & $ 0.54^{+0.36}_{-0.32}$ \\
    & $-0.30^{+0.21}_{-0.23}$ & $ 0.30$         & $1.00^{+0.23}_{-0.21}$ & $0.91^{+0.04}_{-0.05}$ & $ 0.00^{+0.60}$  \\
\hline
\end{tabular}
\end{table}

\begin{figure}
\includegraphics[width=0.8\textwidth]{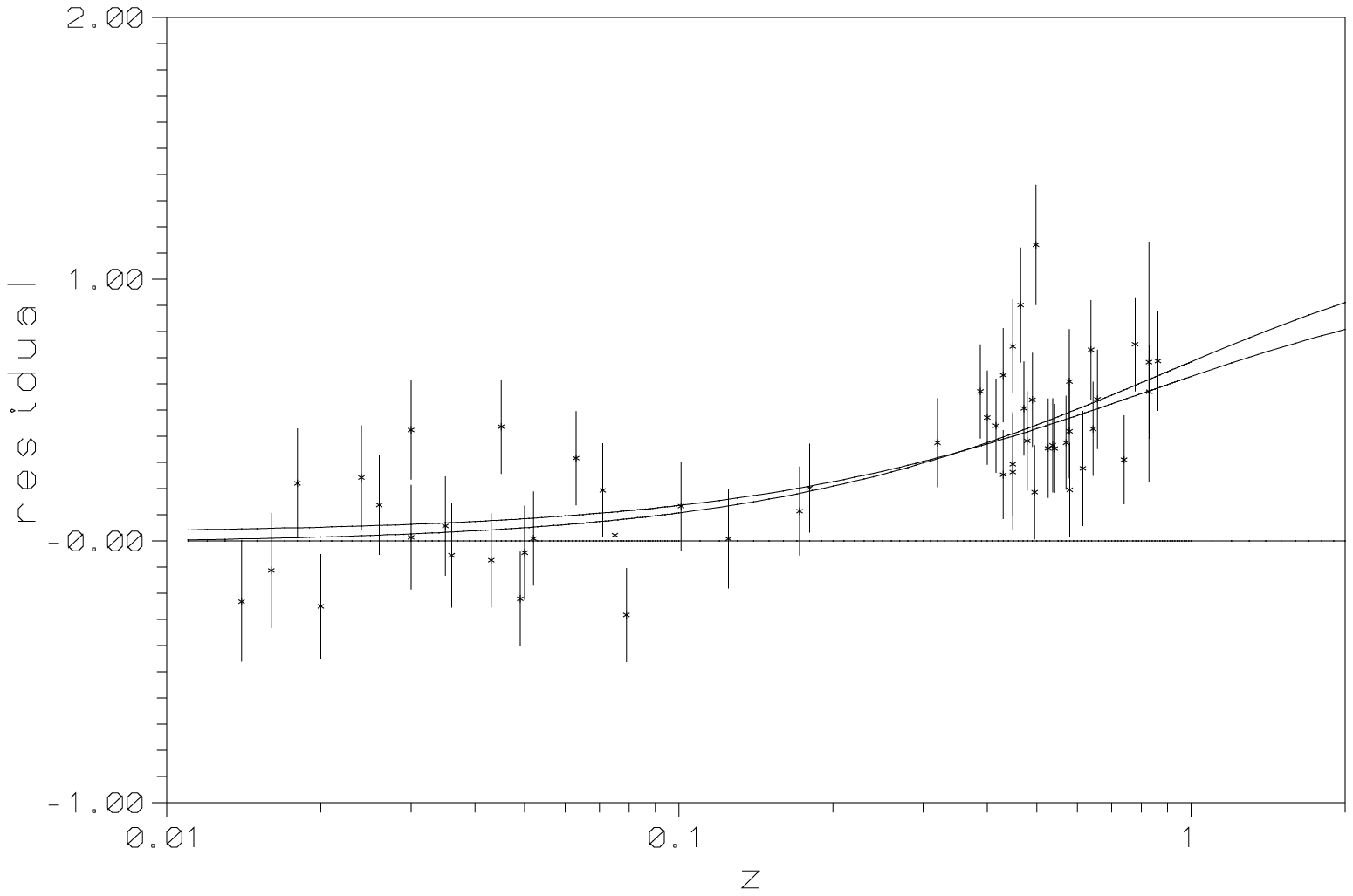}
\caption{Residuals (in mag) between the Einstein-de Sitter model (zero line),
the $\Lambda$CDM model (upper curve) and the best-fitted
Generalized Chaplygin Gas model  with $\Omega_m= 0.3, \Omega_{Ch}= 0.7$
(middle curve), sample K3.}
\label{fig:1}
\end{figure}
 
\begin{figure}
\includegraphics[width=0.8\textwidth]{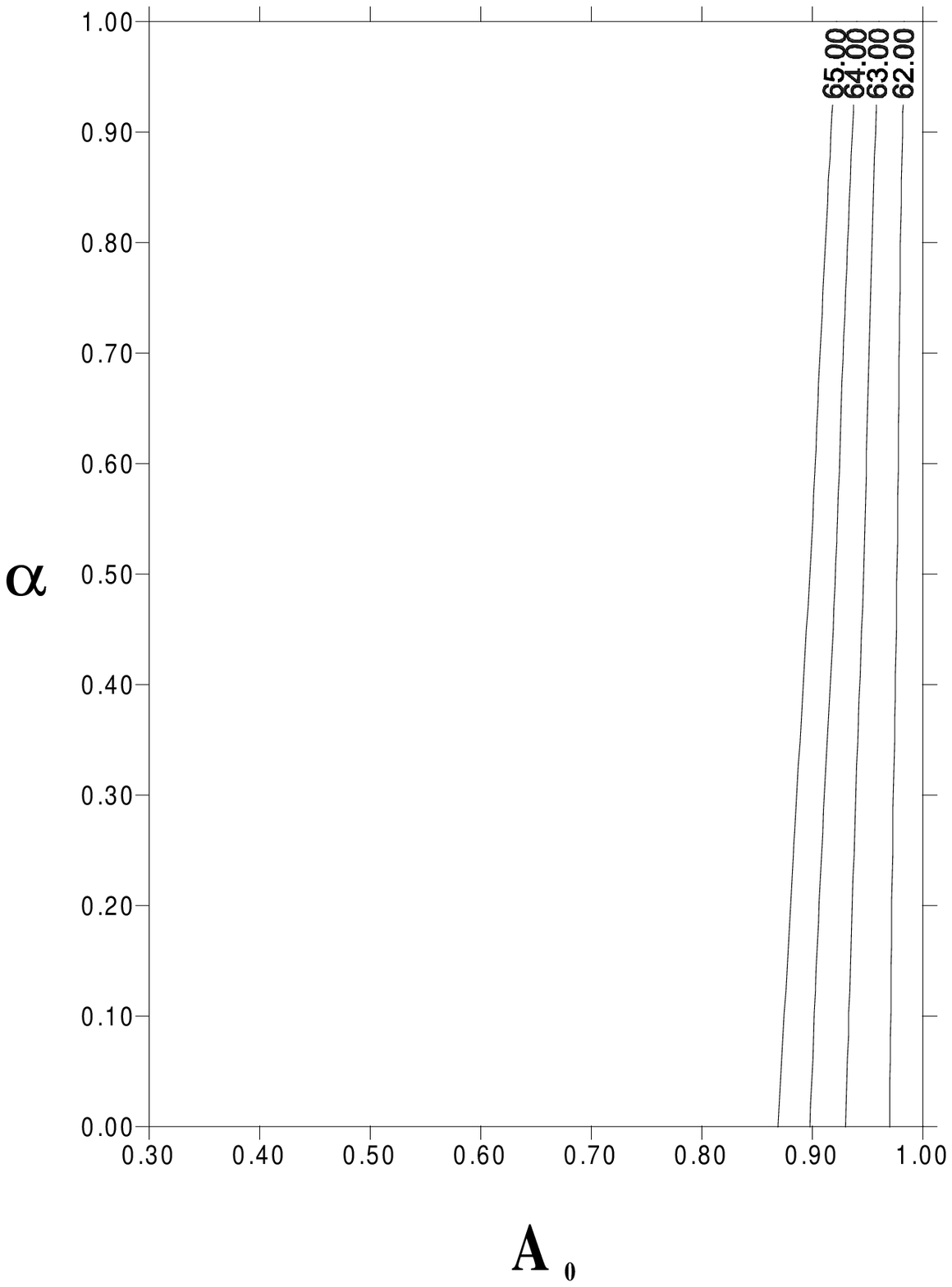}
\caption{Levels of constant $\chi^{2}$ on the plane $(A_0,\alpha)$ for Generalized Chaplygin
Gas model  with $\Omega_m= 0.3, \Omega_{Ch}= 0.7$, sample K3, marginalized
over ${\cal M}$. The figure shows preferred values of $A_0$ and $\alpha$.}
\label{fig:2}
\end{figure}
 
\begin{figure}
\includegraphics[width=0.8\textwidth]{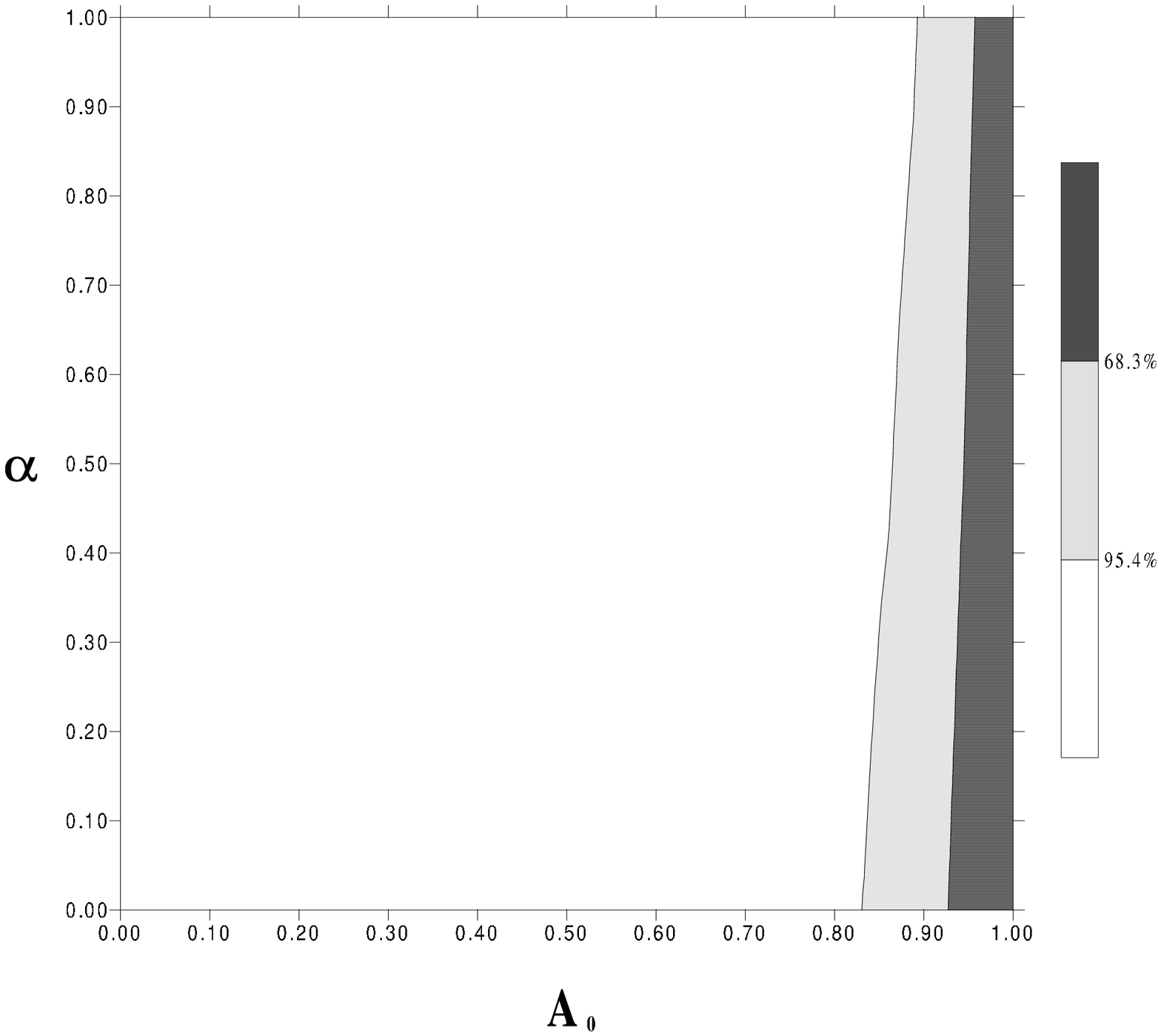}
\caption{Confidence  levels on the plane $(A_0,\alpha)$ for Generalized Chaplygin Gas
model  with $\Omega_m= 0.3, \Omega_{Ch}= 0.7$, sample K3, marginalized over ${\cal M}$.
The figure shows the ellipses of preferred values of $A_0$ and $\alpha$.}
\label{fig:3}
\end{figure}
 
\begin{figure}
\includegraphics[width=0.8\textwidth]{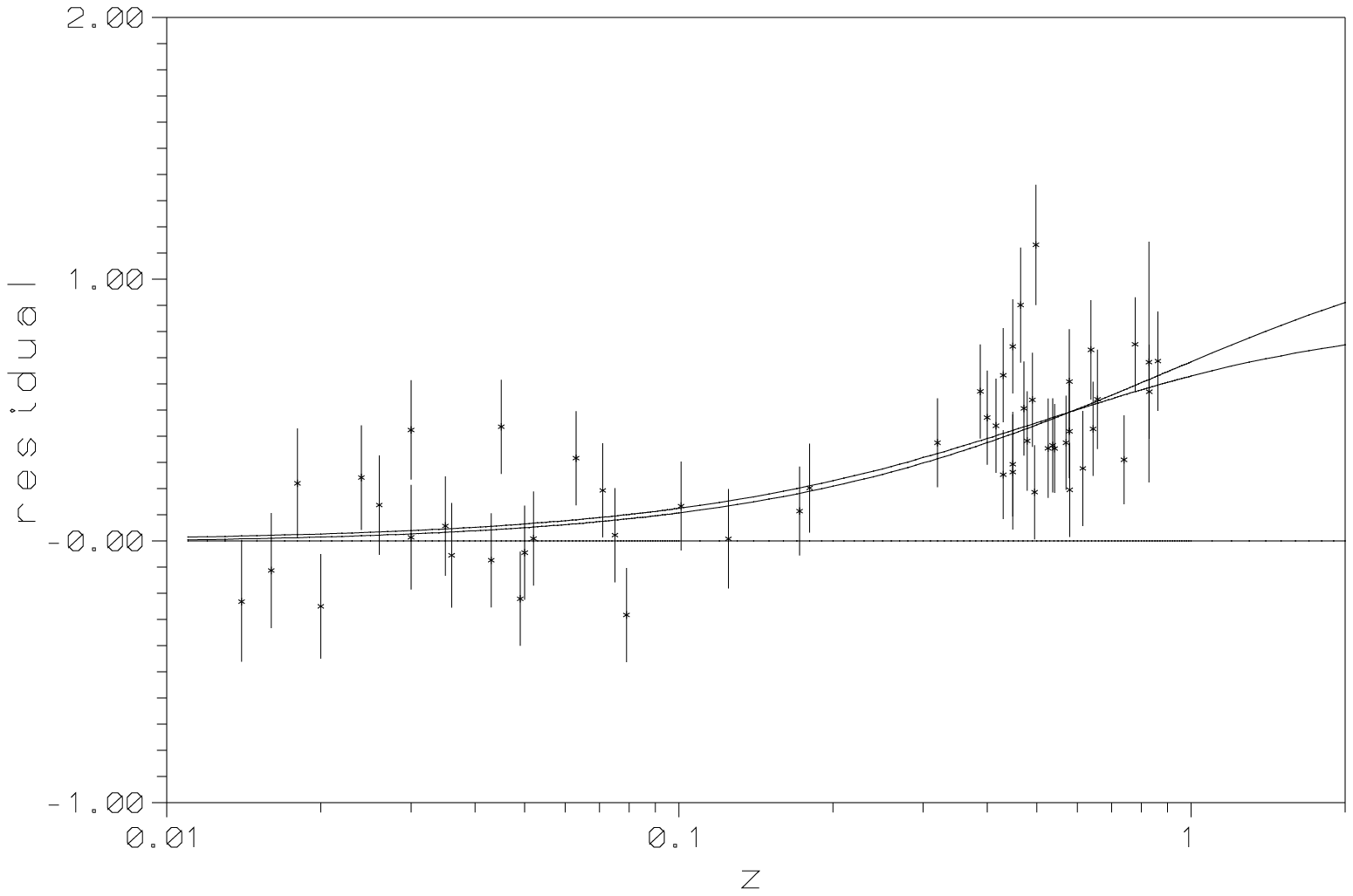}
\caption{Residuals (in mag) between the Einstein-de Sitter model (zero line), the
flat $\Lambda$CDM model (upper curve) and the best-fitted Generalized Chaplygin Gas model
with $\Omega_m= 0.05, \Omega_{Ch}= 0.95$ (middle curve), sample K3.}
\label{fig:4}
\end{figure}
 
\begin{figure}
\includegraphics[width=0.8\textwidth]{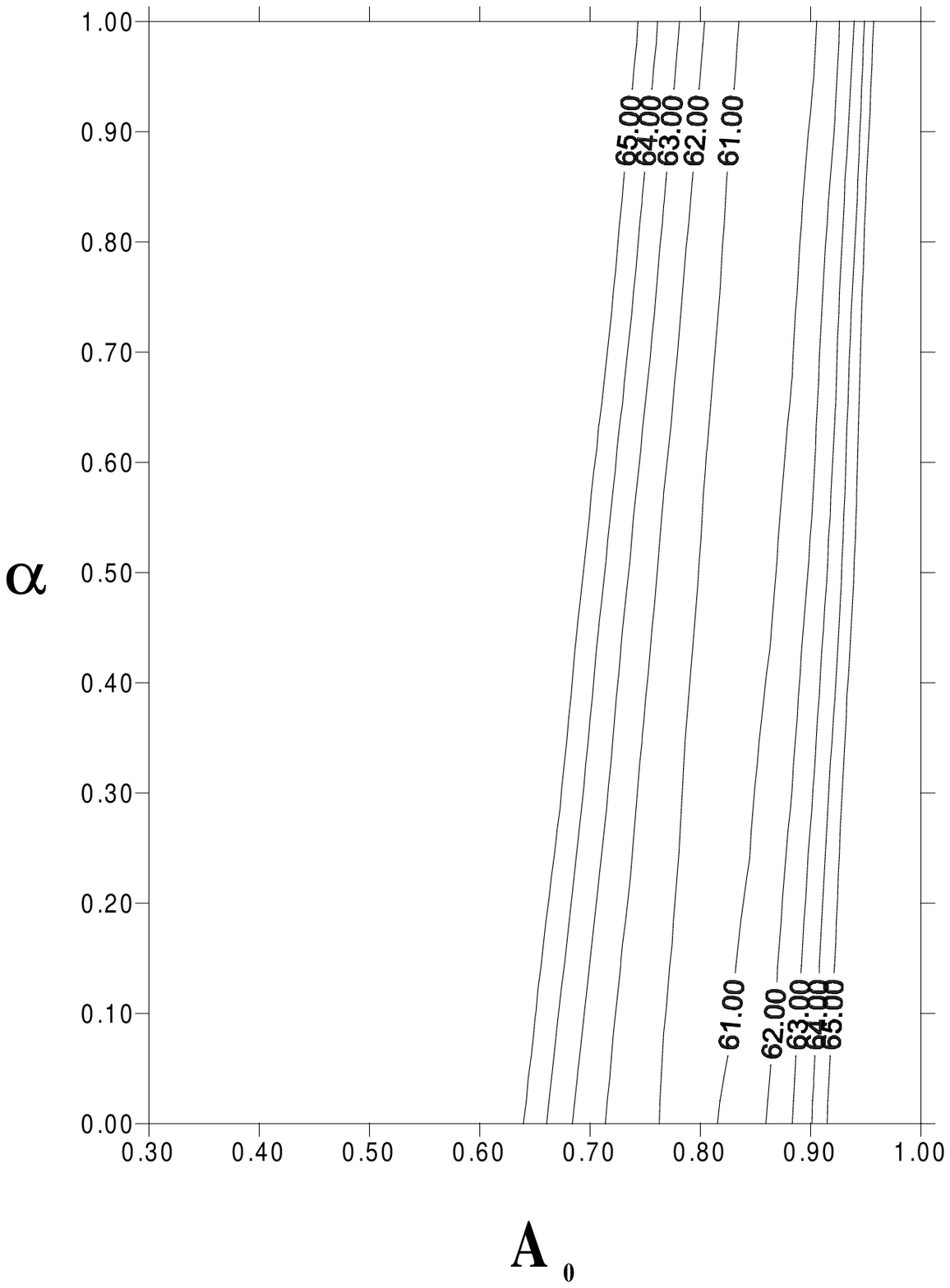}
\caption{Levels of constant $\chi^{2}$ on the plane $(A_0,\alpha)$ for Generalized Chaplygin
Gas model  with $\Omega_m= 0.05, \Omega_{Ch}= 0.95$, sample K3, marginalized over
${\cal M}$. The figure shows preferred values of $A_0$ and $\alpha$.}
\label{fig:5}
\end{figure}
 
\begin{figure}
\includegraphics[width=0.8\textwidth]{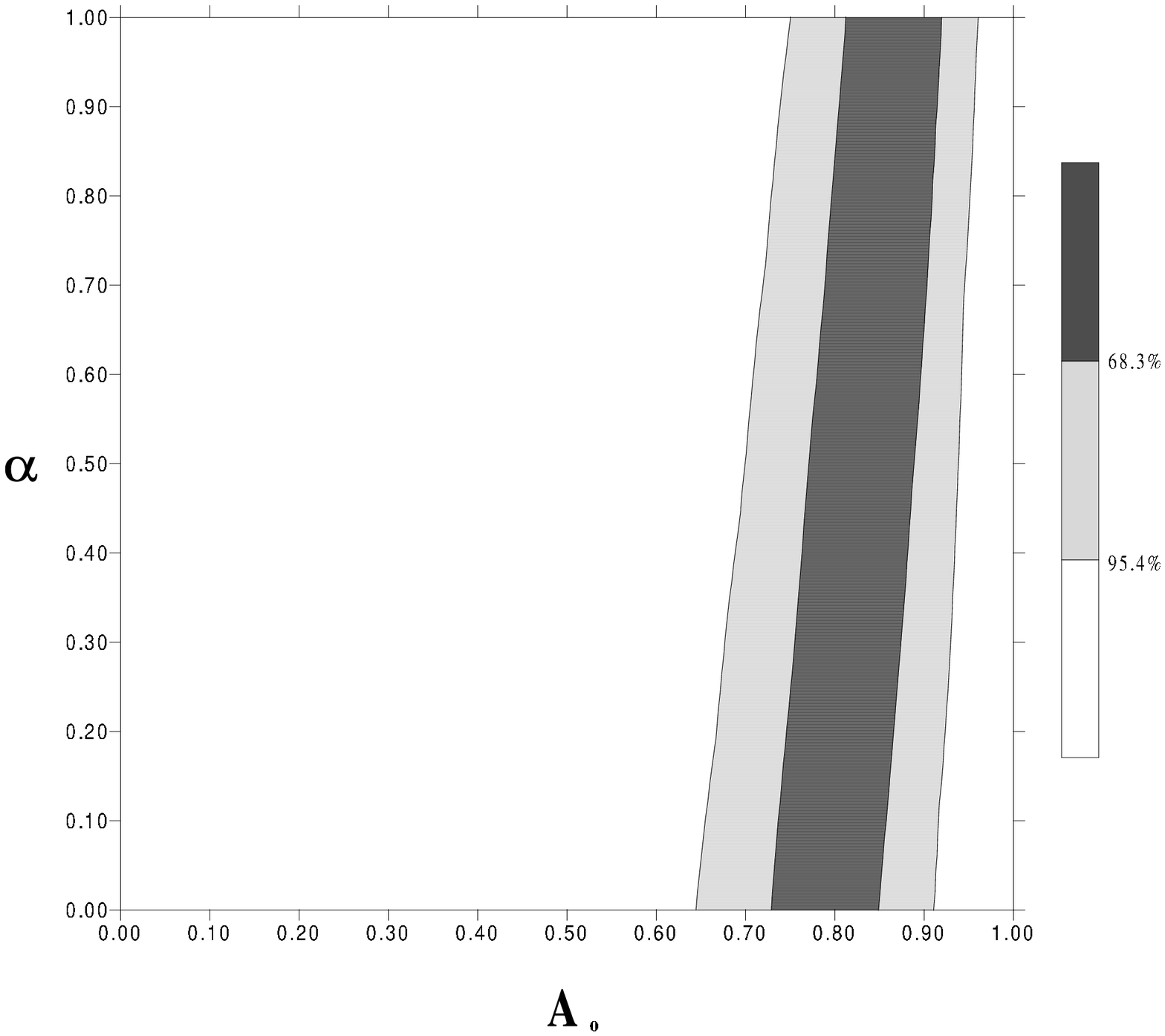}
\caption{Confidence  levels on the plane $(A_0,\alpha)$ for Generalized Chaplygin Gas
model with $\Omega_m= 0.05, \Omega_{Ch}= 0.95$, sample K3, marginalized over ${\cal M}$.
The figure shows the ellipses of preferred values of $A_0$ and $\alpha$.}
\label{fig:6}
\end{figure}
 
\begin{figure}
\includegraphics[width=0.8\textwidth]{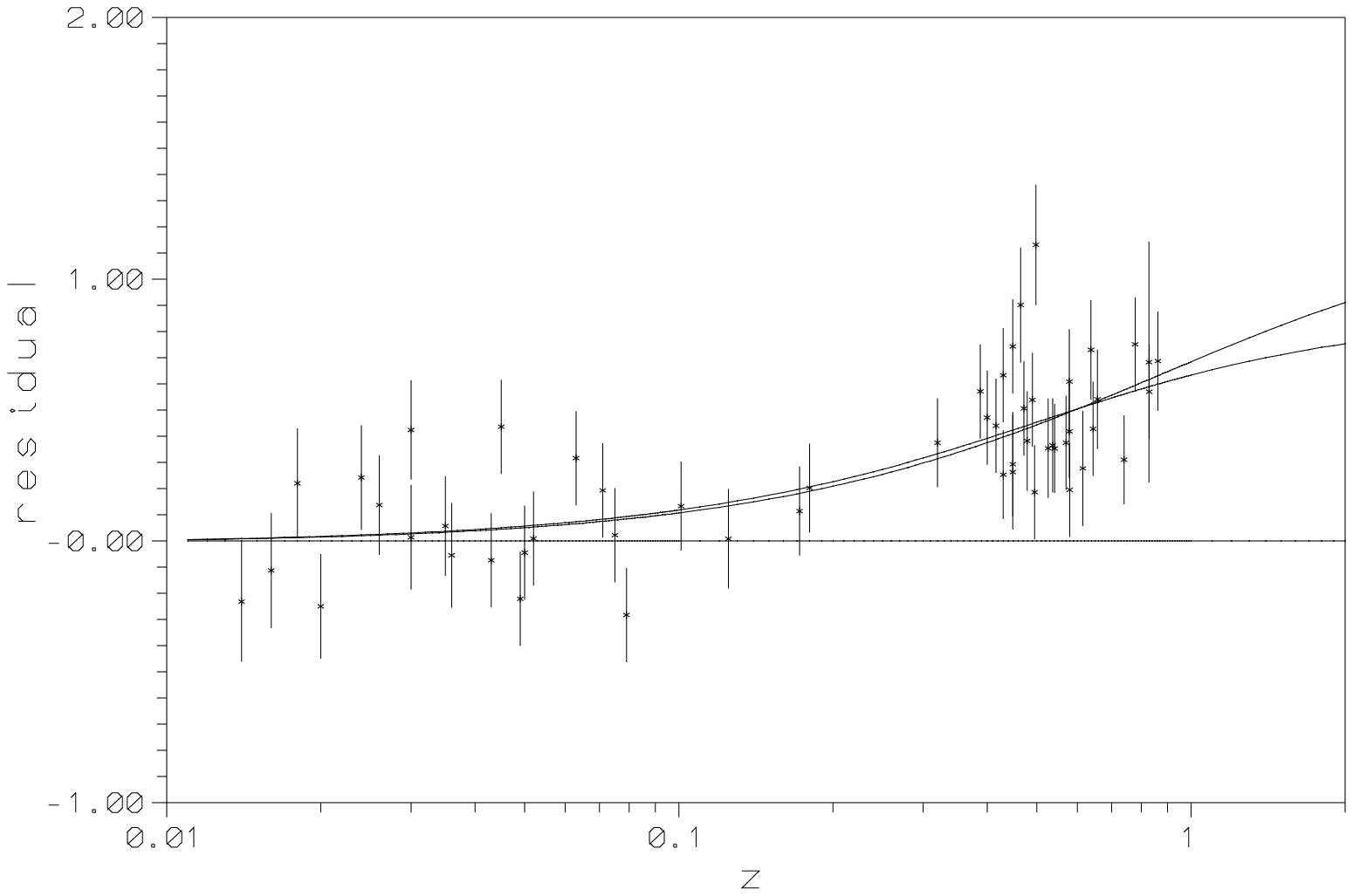}
\caption{Residuals (in mag) between the Einstein-de Sitter model (zero line),
the flat $\Lambda$CDM model (upper curve) and the best-fitted Generalized Chaplygin
Gas model  with $\Omega_m= 0 , \Omega_{Ch}= 1$ (middle curve), sample K3.}
\label{fig:7}
\end{figure}
\begin{figure}
\includegraphics[width=0.8\textwidth]{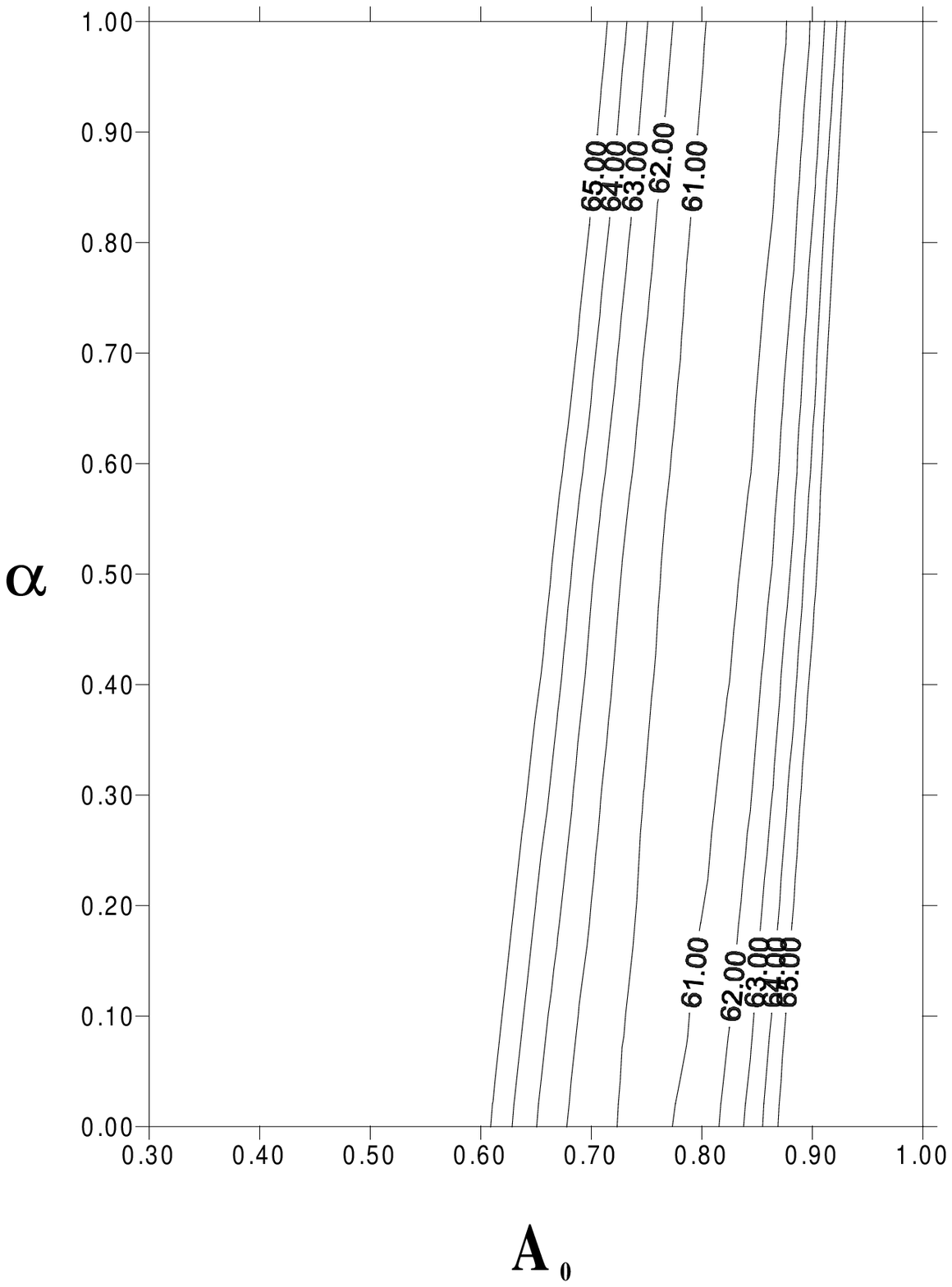}
\caption{Levels of constant $\chi^{2}$ on the plane $(A_0,\alpha)$ for
Generalized Chaplygin Gas model  with $\Omega_m= 0, \Omega_{Ch}= 1$, sample K3,
marginalized over ${\cal M}$. The figure shows preferred values of $A_0$ and $\alpha$.}
\label{fig:8}
\end{figure}
 
\begin{figure}
\includegraphics[width=0.8\textwidth]{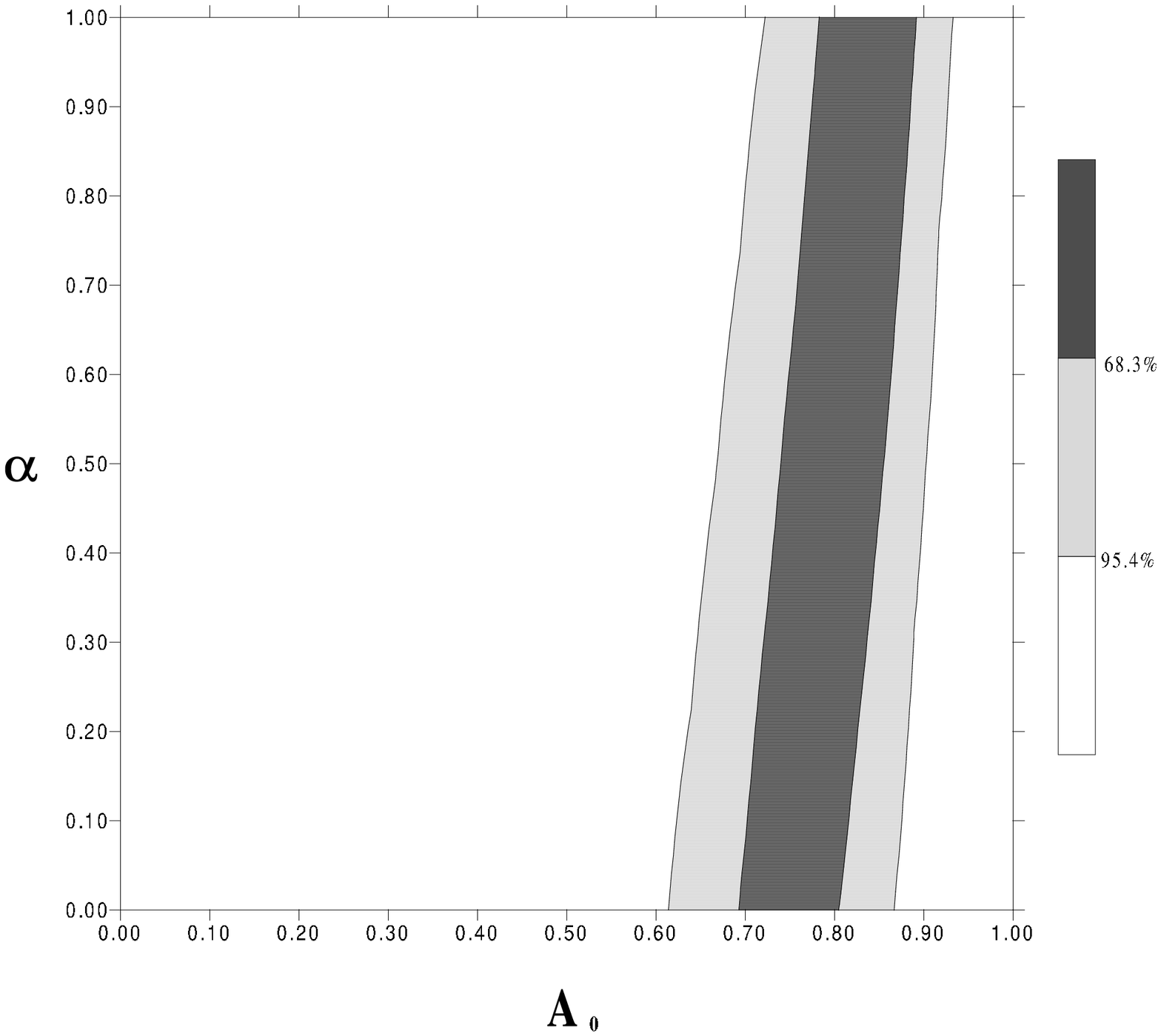}
\caption{Confidence  levels on the plane $(A_0,\alpha)$ for Generalized Chaplygin Gas model
with $\Omega_m= 0, \Omega_{Ch}= 1$, sample K3,  marginalized over ${\cal M}$. The figure
shows the ellipses of preferred values of $A_0$ and $\alpha$.}
\label{fig:9}
\end{figure}
 
\begin{figure}
\includegraphics[width=0.8\textwidth]{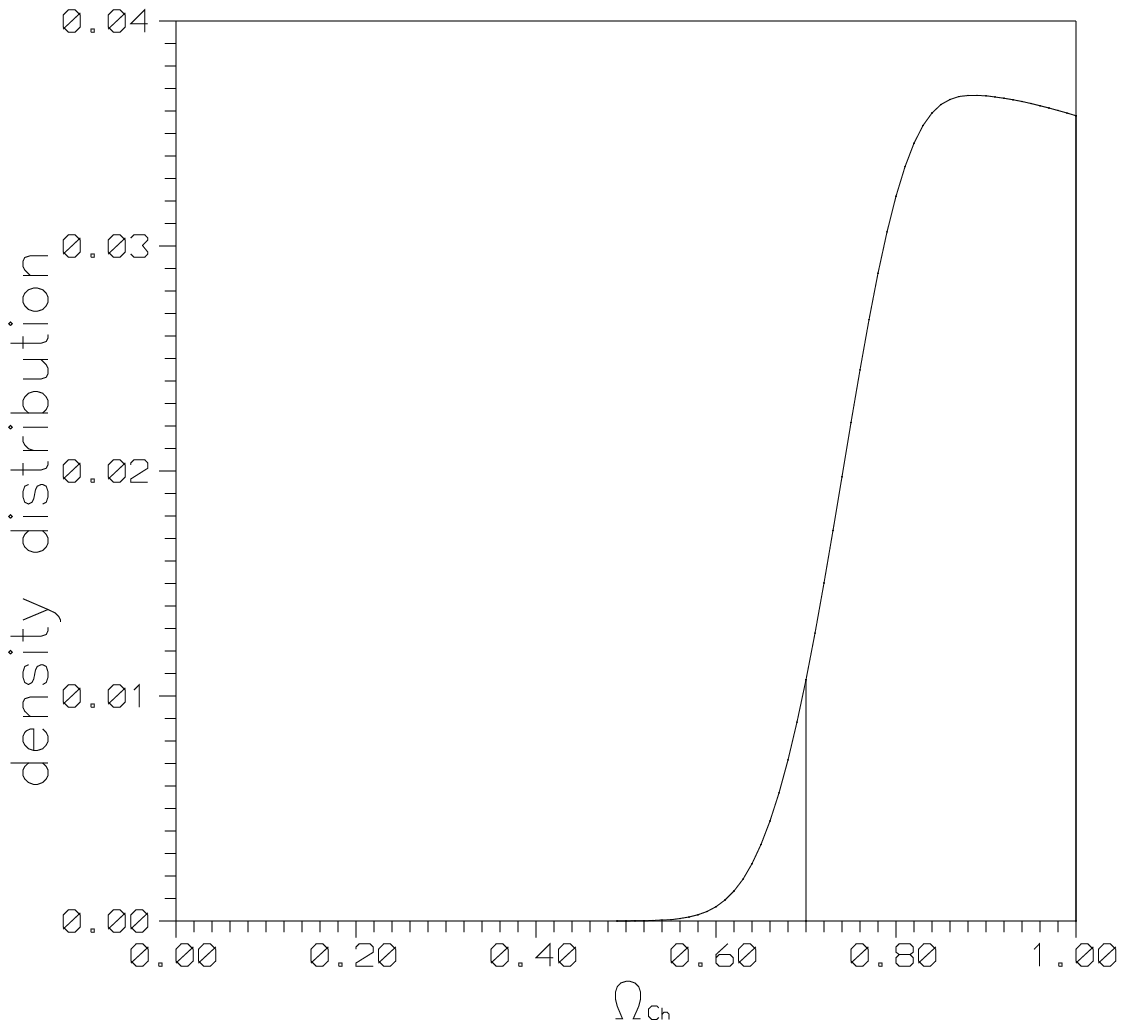}
\caption{The density distribution (one dimensional PDF) for $\Omega_{Ch}$ obtained from
sample K3 by marginalization over remaining parameters of the model. We obtain the
limit $\Omega_{Ch} > 0.70$ at the confidence level $95.4 \%$.}
\label{fig:10}
\end{figure}

\section{Cosmological model}
 
Einstein equations for the Friedman-Robertson-Walker model with hydrodynamical
energy-momentum tensor
$T_{\mu \nu} = (\rho + p) u_{\mu} u_{\nu} - p g_{\mu \nu}$ read:
\begin{eqnarray}
\left( \frac{\dot a}{a} \right)^2 &= & \frac{8 \pi G \rho}{3} - \frac{k}{a^2(t)} \label{first}\\
\frac{{\ddot a(t)}}{a} &= & - \frac{4 \pi G}{3}(\rho + 3p) \label{Raychaud}
\end{eqnarray}
Let us assume that matter content of the Universe consists of pressure-less gas
with energy density $\rho_m$ representing baryonic plus cold dark matter (CDM)
and the Generalized Chaplygin Gas with the equation  of state
\begin{equation}
\label{EOS}
p_{Ch} = - \frac{A}{{\rho_{Ch}}^{\alpha}}
\end{equation}
representing the dark energy responsible for the acceleration of the Universe.
If one further makes an assumption that these two components do not interact, then the energy
conservation equation
\begin{equation}
\label{conservation}
{\dot \rho} + 3 H (p + \rho) =  0
\end{equation}
where $H =  {\dot a}/a$ is the Hubble function, can be integrated separately for matter
and Chaplygin gas leading to well known result
$\rho_m =  \rho_{m,0} a^{-3}$ and
(see also Bento, Bertolami \& Sen 2002 or Carturan \& Finelli 2002)
\begin{equation}
\label{Chap}
\rho_{Ch} =  \left( A + \frac{B}{a^{3(1 + \alpha)}} \right)^{\frac{1}{1+ \alpha}}
\end{equation}
 
The physical interpretation of, so far arbitrary, constants $A$ and $B$ is the following.
Adopting usual convention that current value of the scale factor $a_0$ is equal to 1, one
can see that $\rho_{Ch,0} =  ( A + B)^{\frac{1}{1+ \alpha}}$ represents the current
energy density of the Chaplygin gas. Calculating the adiabatic speed of sound squared
for the Chaplygin gas
$$c_s^2 =  \frac{\partial p_{Ch}}{\partial \rho_{Ch}}
=  \frac{\alpha A}{\rho^{1 + \alpha}} =   \frac{\alpha A}{A + \frac{B}{a^{3(1 + \alpha)}}}$$
it is easy to confirm that the current value of $c_s^2$ is equal to
$c_{s,0}^2 = \frac{\alpha A}{A+B}$.
Hence the constants $A$ and $B$ can be expressed as combinations of quantities having
well defined physical meaning.
 
Our further task will be to confront the Chaplygin gas model with SNIa data and for
this purpose we have to calculate the luminosity distance in our model
\begin{equation}
\label{luminosity_dist}
d_L(z) =  (1+z) \frac{c}{H_0} \frac{1}{\sqrt{|\Omega_k|}}
{\cal F} \left( H_0 \sqrt{|\Omega_k|} \int_0^z \frac{d z'}{H(z')} \right)
\end{equation}
where $\Omega_k =  - \frac{k}{H_0^2}$ and
\begin{eqnarray}
{\cal F}(x) =  sinh(x) \;\;\; &for& \;\;\; k<0  \\
{\cal F}(x) =  x \;\;\; &for& \;\;\; k= 0  \\
{\cal F}(x) =  sin(x) \;\;\; &for& \;\;\; k>0
\end{eqnarray}
 
The Friedman equation (\ref{first}) can be rearranged to the form giving
explicitly the Hubble function $H(z)= {\dot a}/a$
\begin{equation}
\label{Hubble}
H(z)^2 =  H_0^2 \left[ \Omega_{m} (1+z)^3 + \Omega_{Ch}
\left(A_0 + (1 - A_0)(1+z)^{3(1+ \alpha)} \right)^{\frac{1}
{1+\alpha}} + \Omega_{k} (1+z)^2 \right]
\end{equation}
where the quantities $\Omega_i$, $i= m,Ch,k$ represent fractions of critical density currently
contained in energy densities of respective components and
$\Omega_m + \Omega_{Ch} + \Omega_k =  1$. For the transparency of further formulae we have
also denoted $A_0 =  A /(A+B)$.
 
Finally the luminosity distance reads:
\begin{eqnarray}
&&d_L(z) =  (1+z) \frac{c}{H_0} \frac{1}{\sqrt{|\Omega_k|}}\nonumber\\
&&{\cal F} \left( \sqrt{|\Omega_k|} \int_0^z \frac{d z'}{\sqrt{\Omega_m (1+z)^3 + \Omega_{Ch}
\left(A_0 + (1 - A_0)(1+z)^{3(1+ \alpha)} \right) ^\frac{1} {1+\alpha}+ \Omega_k (1+z)^2 }}
\right)\nonumber\\
\label{LD}
\end{eqnarray}
 
The formula (\ref{LD}) is the most general one in the framework of
Friedman-Robertson-Walker cosmology with Generalized Chaplygin Gas.
Please note, that this model propose a unified macroscopic
(phenomenological) description of both dark energy and dark matter.
 
Further in this paper we will mostly use its version restricted to flat
model $k= 0$ (the exception will be when we relax flat prior)
since the evidence for this case is very strong in the light
of current CMBR data. Therefore while talking about model testing we actually
 mean the estimation of $\alpha$ and $A_0$ parameters for the
best fitted flat FRW cosmological model filled with Generalized Chaplygin Gas.
 
To proceeded with fitting the SNIa data we need the magnitude-redshift relation
\begin{equation}
\label{m-z}
m(z,{\cal M}, \Omega_m, \Omega_{Ch};A_0, \alpha) =
{\cal M} + 5 \log_{10} D_L (z, \Omega_m, \Omega_{Ch};A_0, \alpha)
\end{equation}
where:
$$ D_L (z, \Omega_m, \Omega_{Ch};A_0, \alpha) =
H_0 d_L (z, H_0, \Omega_m, \Omega_{Ch};A_0, \alpha) $$ is the luminosity distance with
$H_0$ factored out so that marginalization over the intercept
\begin{equation}
\label{intercept}
{\cal M} =  M - 5 \log_{10} H_0 +25
\end{equation}
leads actually to joint marginalization over $H_0$ and $M$ ($M$ being the
absolute magnitude of SNIa).
 
Then we can obtain the best fitted model minimalizing the $\chi^2$ function
$$ \chi^2 =  \sum_i \frac{(m_i^{Ch} - m_i^{obs})^2}{\sigma_i^2} $$ where the sum is over
the SNIa sample and $\sigma_i$ denote the (full) statistical error of magnitude determination.
This is illustrated by figures (Fig.1 - Fig.9) of residuals (with respect to Einstein-de Sitter model)
and $\chi^2$ levels in the $(A_0, \alpha)$ plane. One of the advantages of residual plots is
that the intercept of the $m - z$ curve gets cancelled. The assumption that the intercept
is the same for different cosmological models is legitimate since ${\cal M}$ is actually
determined from the low-redshift part of the Hubble diagram which should be linear in all
realistic cosmologies. From the full Perlmutter's sample (see below) we have obtained
${\cal M} =  -3.39$ which is in a very good agreement with the values reported in the
literature (Perlmutter et al. 1999, Riess et al. 1999). For other samples see discussion below.
 
The best-fit values alone are not relevant if not supplemented with the confidence
levels for the parameters. Therefore, we performed the estimation of model parameters using
the minimization procedure, based on the likelihood function. We assumed that
supernovae measurements came with uncorrelated Gaussian errors and in this case the
likelihood function ${\cal L}$ could be determined from chi-square statistic
${\cal L} \propto \exp{(-\chi^2/2)}$ (Perlmutter et al. 1999, Riess et al. 1999).
 
Therefore we supplement our analysis with confidence intervals in the $(A_0, \alpha)$ plane by
calculating the marginal probability density functions
$$ {\cal P}(A_0, \alpha) \propto \int
\exp{(-\chi^2(\Omega_m, \Omega_{Ch}, A_0, \alpha, {\cal M})/2)} d{\cal M} $$
with $\Omega_m, \Omega_{Ch}$ fixed ($\Omega_m$ =  0.0, 0.05, 0.3 ) and
$$ {\cal P}(A_0, \alpha) \propto \int \exp{(-\chi^2(\Omega_m, \Omega_{Ch}, A_0, \alpha,
{\cal M})/2)} d{\Omega_m} $$
with ${\cal M}$ fixed (${\cal M} =  -3.39$) respectively (proportionality sign means
equal up to the normalization constant). In order to complete the picture we have also
derived one-dimensional probability distribution functions for $\Omega_{Ch}$ obtained from
joint marginalization over $\alpha$ and $A_0$. The maximum value of such PDF informs us about
the most probable value of $\Omega_{Ch}$ (supported by supernovae data) within the full class
of Generalized Chaplygin Gas models.
 
\begin{figure}
\includegraphics[width=0.8\textwidth]{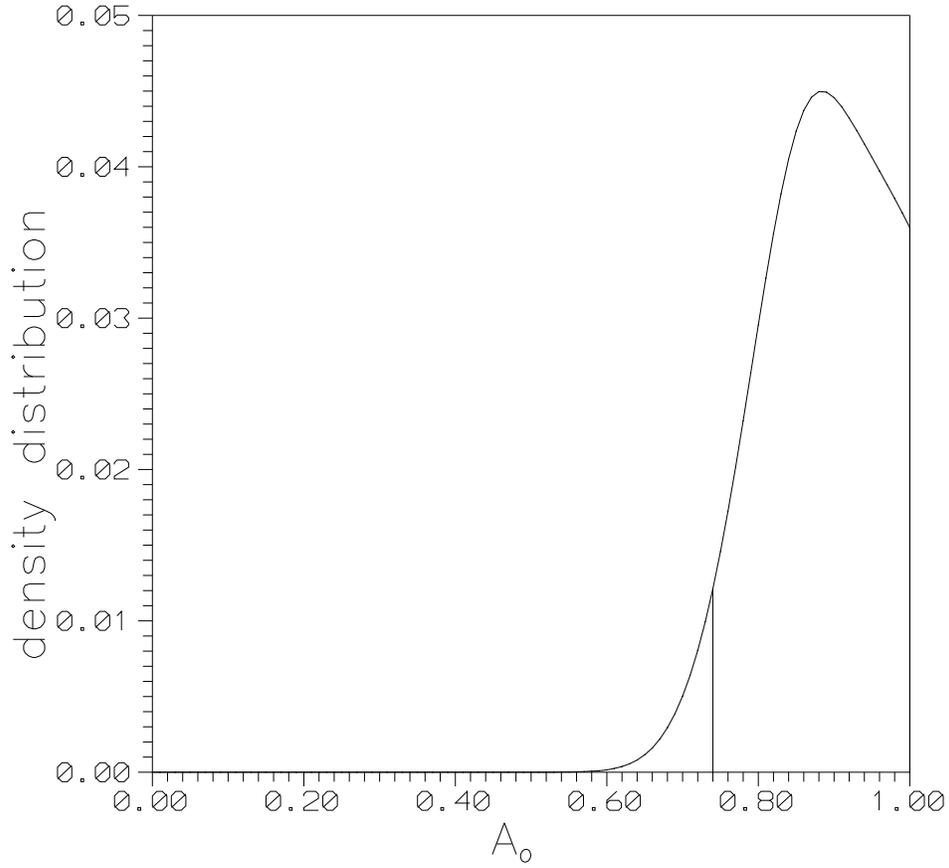}
\caption{The density distribution (one dimensional PDF) for $A_0$ obtained from sample
K3 by marginalization over remaining parameters of the model. We obtain the limit
$A_0 > 0.74$ at the confidence level $95.4 \%$.}
\label{fig:11}
\end{figure}
 
\begin{figure}
\includegraphics[width=0.8\textwidth]{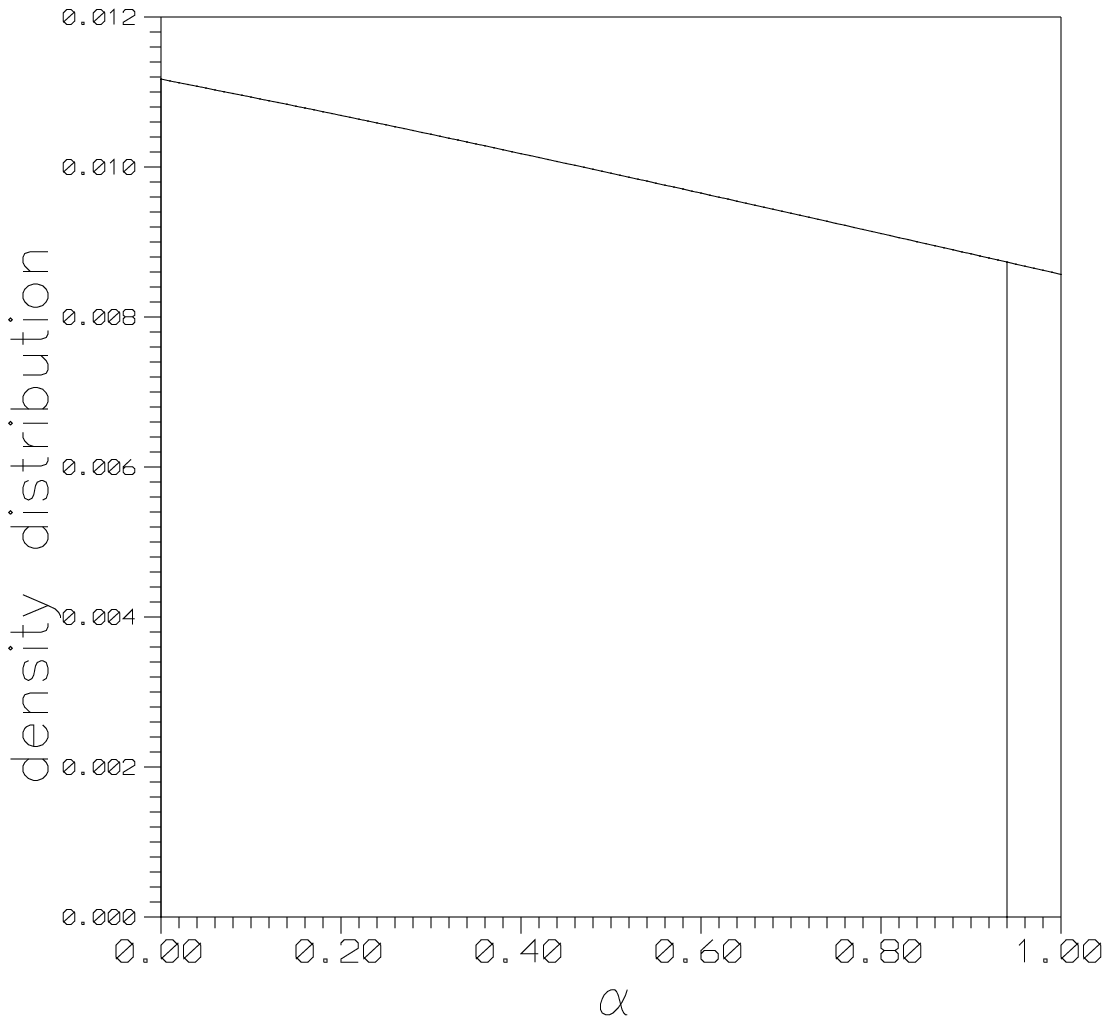}
\caption{The density distribution (one dimensional PDF) for $\alpha$ obtained from sample
K3 by marginalization over remaining parameters of the model. We obtain the limit
$\alpha < 0.94$ at the confidence level $95.4 \%$ .}
\label{fig:12}
\end{figure}

\begin{figure}
\includegraphics[width=0.8\textwidth]{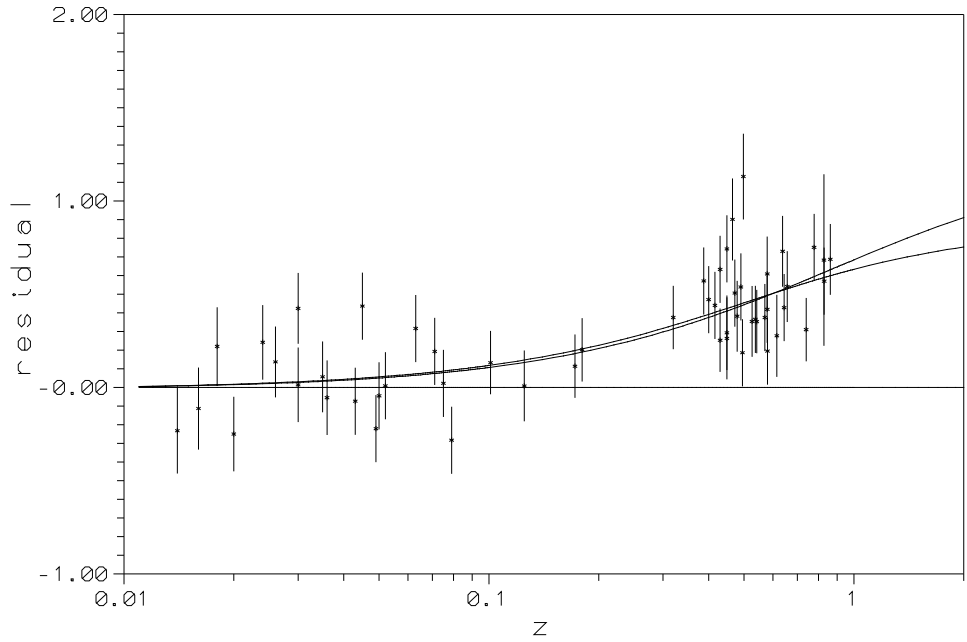}
\caption{Residuals (in mag) between the Einstein-de Sitter model (zero line), the flat
$\Lambda$CDM model ( upper curve) and the best-fitted Generalized Chaplygin Gas model
(without any prior assumptions on $\Omega_m$) (middle curve), sample K3.}
\label{fig:13}
\end{figure}
 
\begin{figure}
\includegraphics[width=0.8\textwidth]{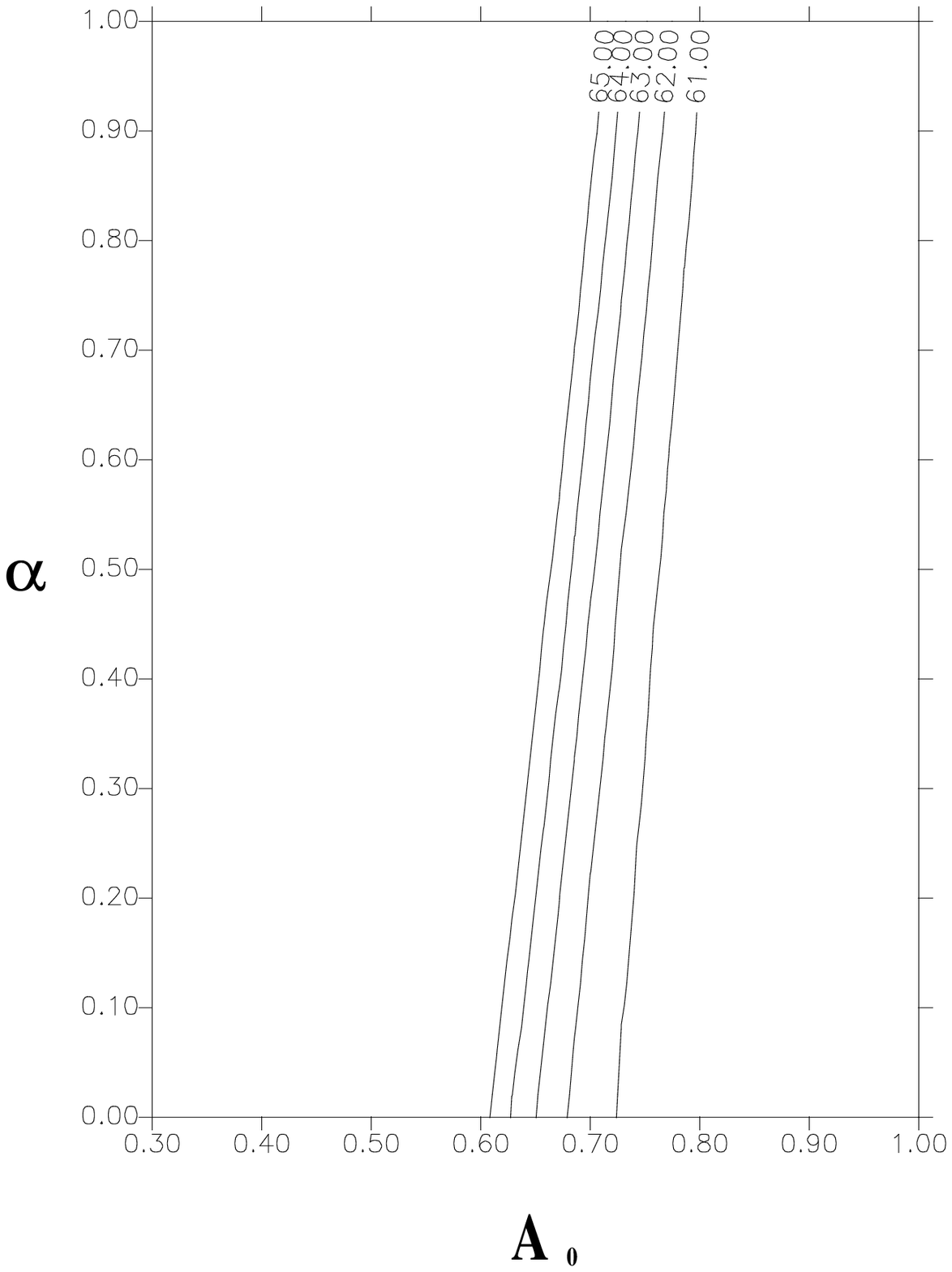}
\caption{Levels of constant $\chi^{2}$ on the plane $(A_0,\alpha)$ for Generalized
Chaplygin Gas model, sample K3, marginalized over ${\cal M}$, and $\Omega_m$.
The figure shows preferred values of $A_0$ and $\alpha$.}
\label{fig:14}
\end{figure}
 
\begin{figure}
\includegraphics[width=0.8\textwidth]{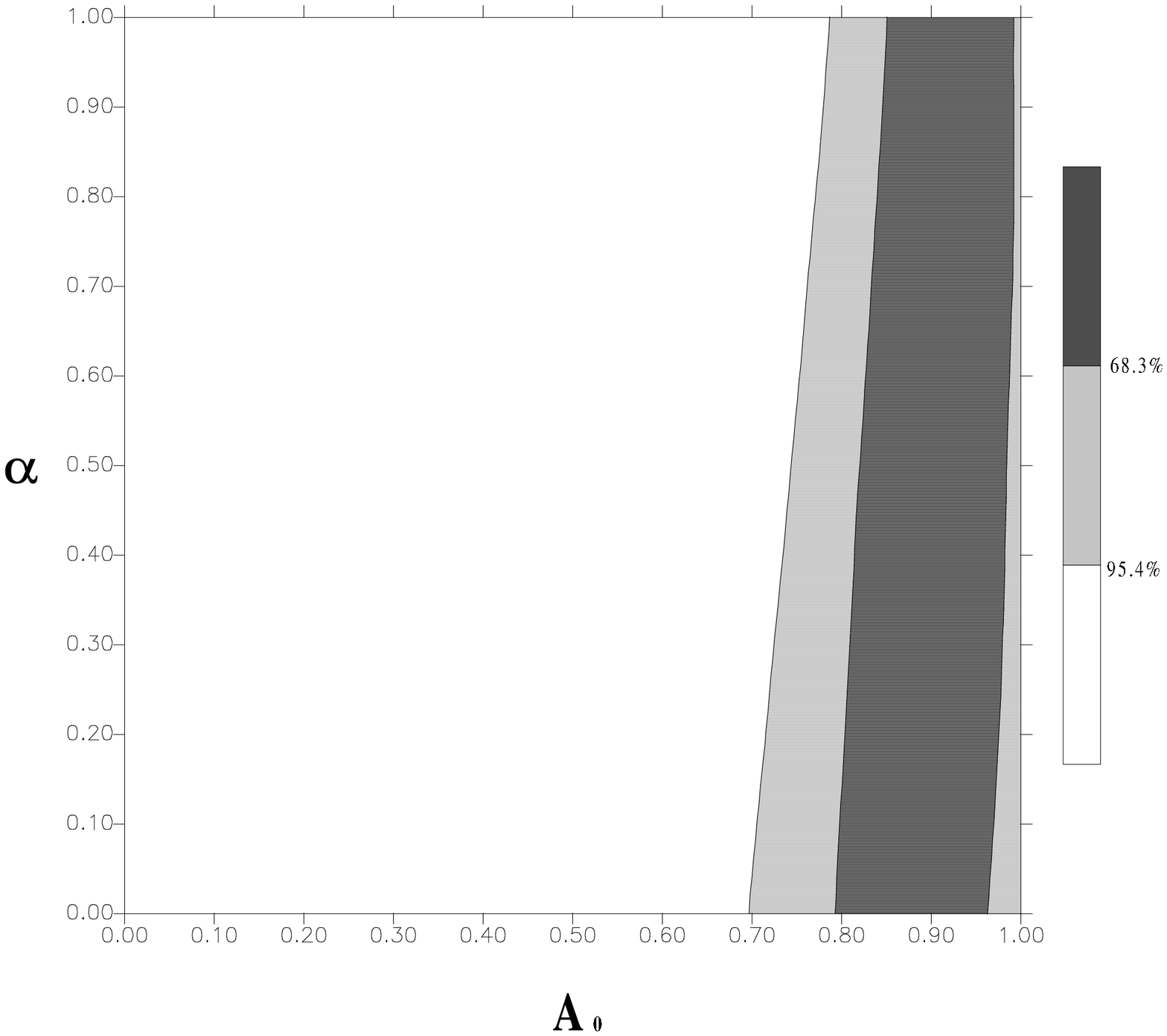}
\caption{Confidence  levels on the plane $(A_0,\alpha)$ for Generalized Chaplygin Gas model,
sample K3, marginalized over ${\cal M}$, and $\Omega_m$.
The figure shows the ellipses of preferred values of $A_0$ and $\alpha$.}
\label{fig:15}
\end{figure}
 
\begin{figure}
\includegraphics[width=0.8\textwidth]{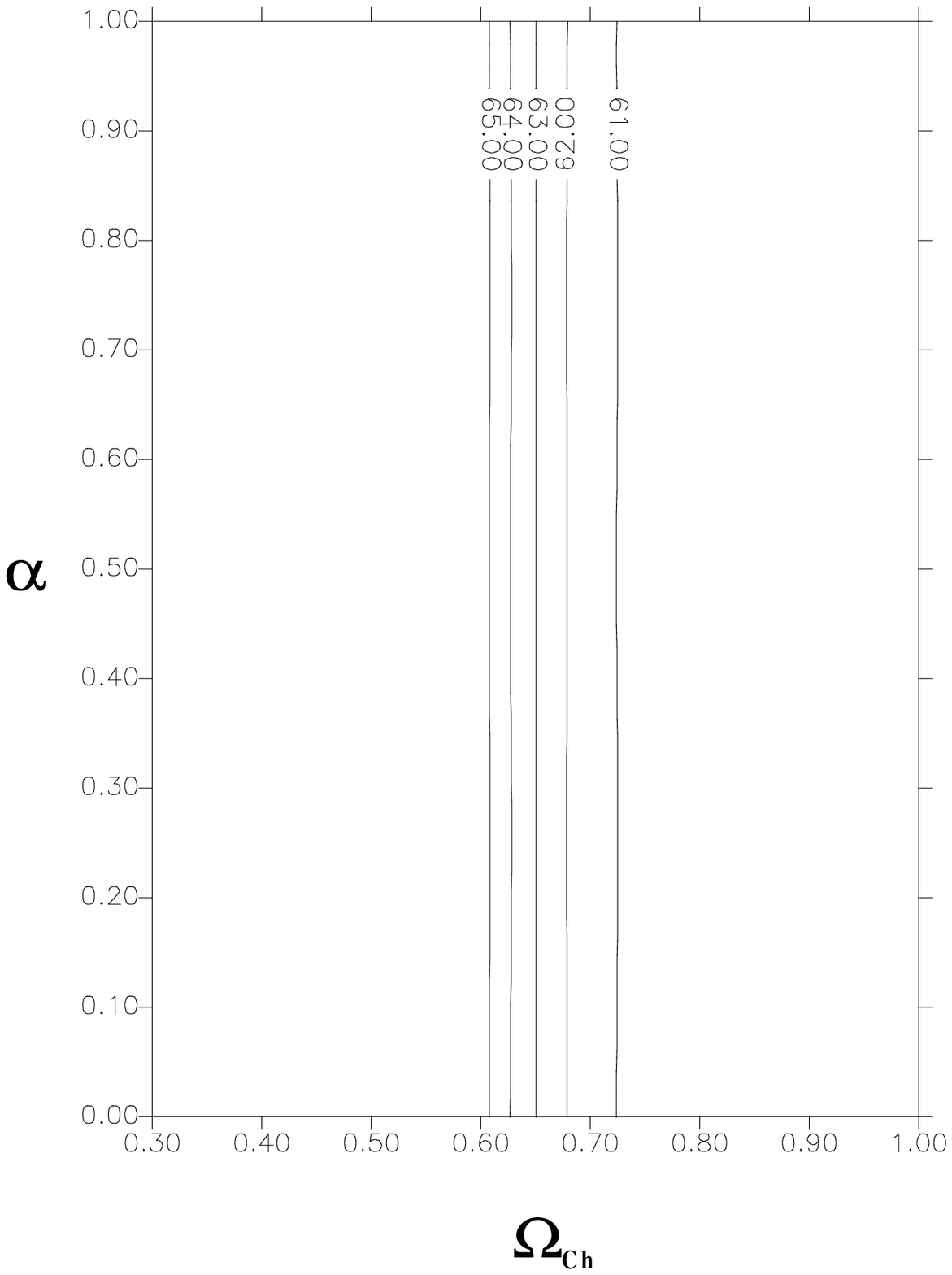}
\caption{Levels of constant $\chi^{2}$ on the plane $(\Omega_m,\alpha)$ for
Generalized Chaplygin Gas model, sample K3, marginalized over ${\cal M}$, and $A_0$.
The figure shows preferred values of $\Omega_m,$  and $\alpha$.} \label{fig:16}
\end{figure}
 
\begin{figure}
\includegraphics[width=0.8\textwidth]{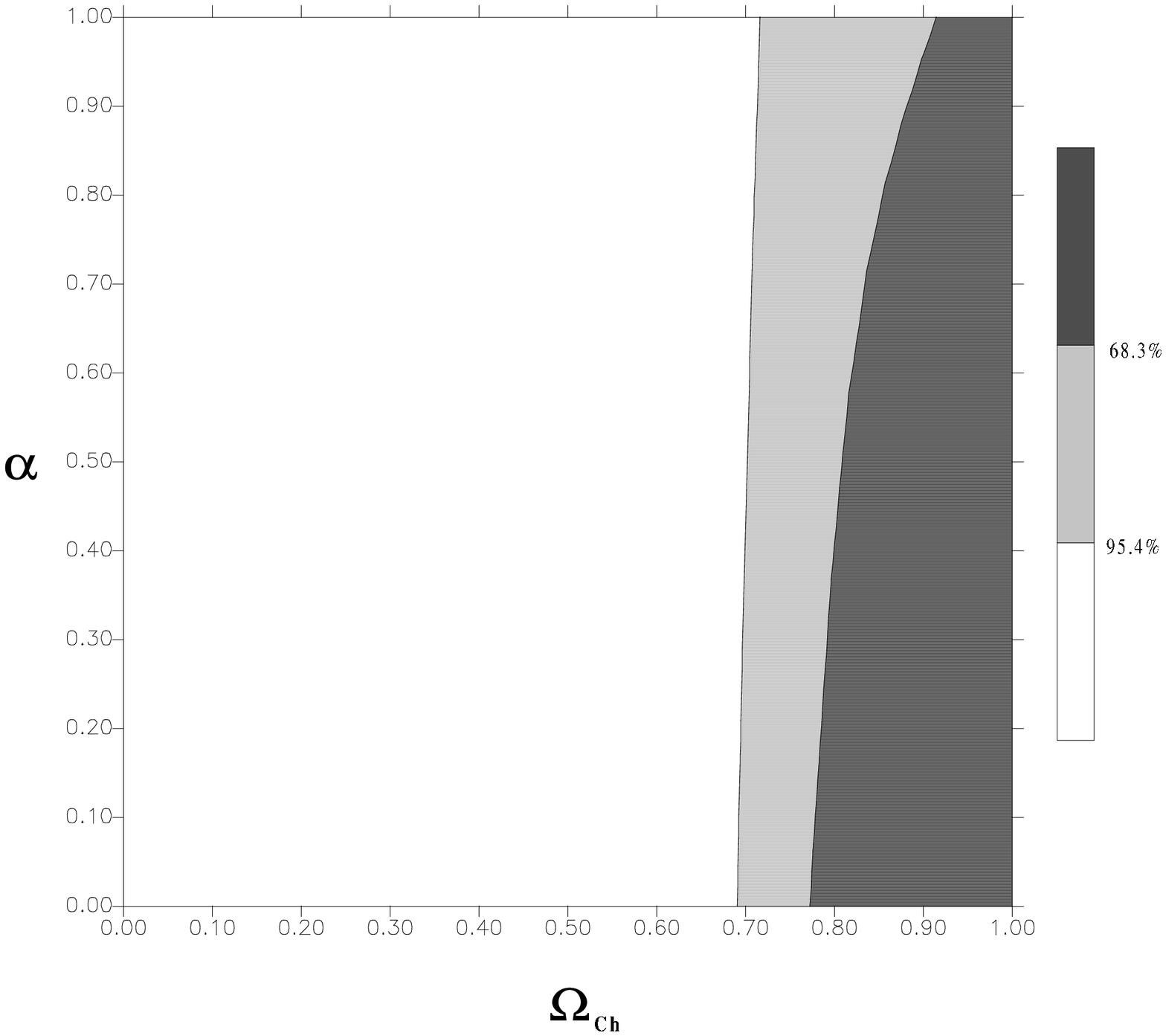}
\caption{Confidence  levels on the plane $(\Omega_m, \alpha)$ for Generalized Chaplygin Gas
model, sample K3, marginalized over $\mathcal{M}$, and $A_0$.
The figure shows the ellipses of preferred values of $\Omega_m$,and $\alpha$.}
\label{fig:17}
\end{figure}
 
\begin{figure}
\includegraphics[width=0.8\textwidth]{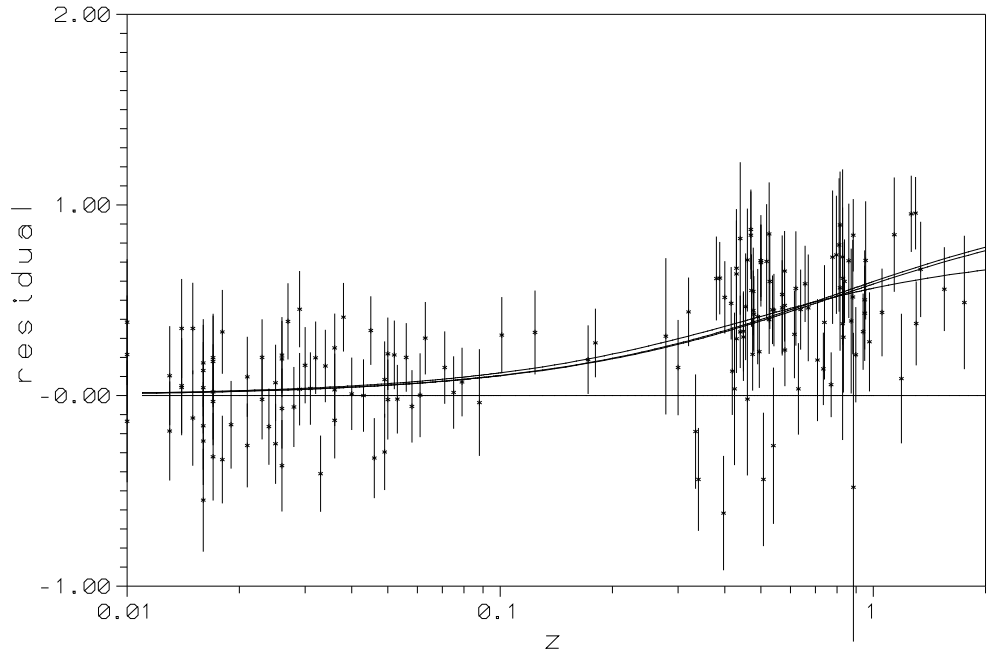}
\caption{Residuals (in mag) between the Einstein-de Sitter model (zero line),
flat $\Lambda$CDM model (two upper curves: for SNIA with $z<1$ --- curve located higher
and for all supernovae Ia belonging to the sample --- curve located lower)
and the best-fitted Generalized Chaplygin Gas model
(without any prior assumptions on $\Omega_m$) (middle curve), GOLD sample.}
\label{fig:18}
\end{figure}
 
\begin{figure}
\includegraphics[width=0.8\textwidth]{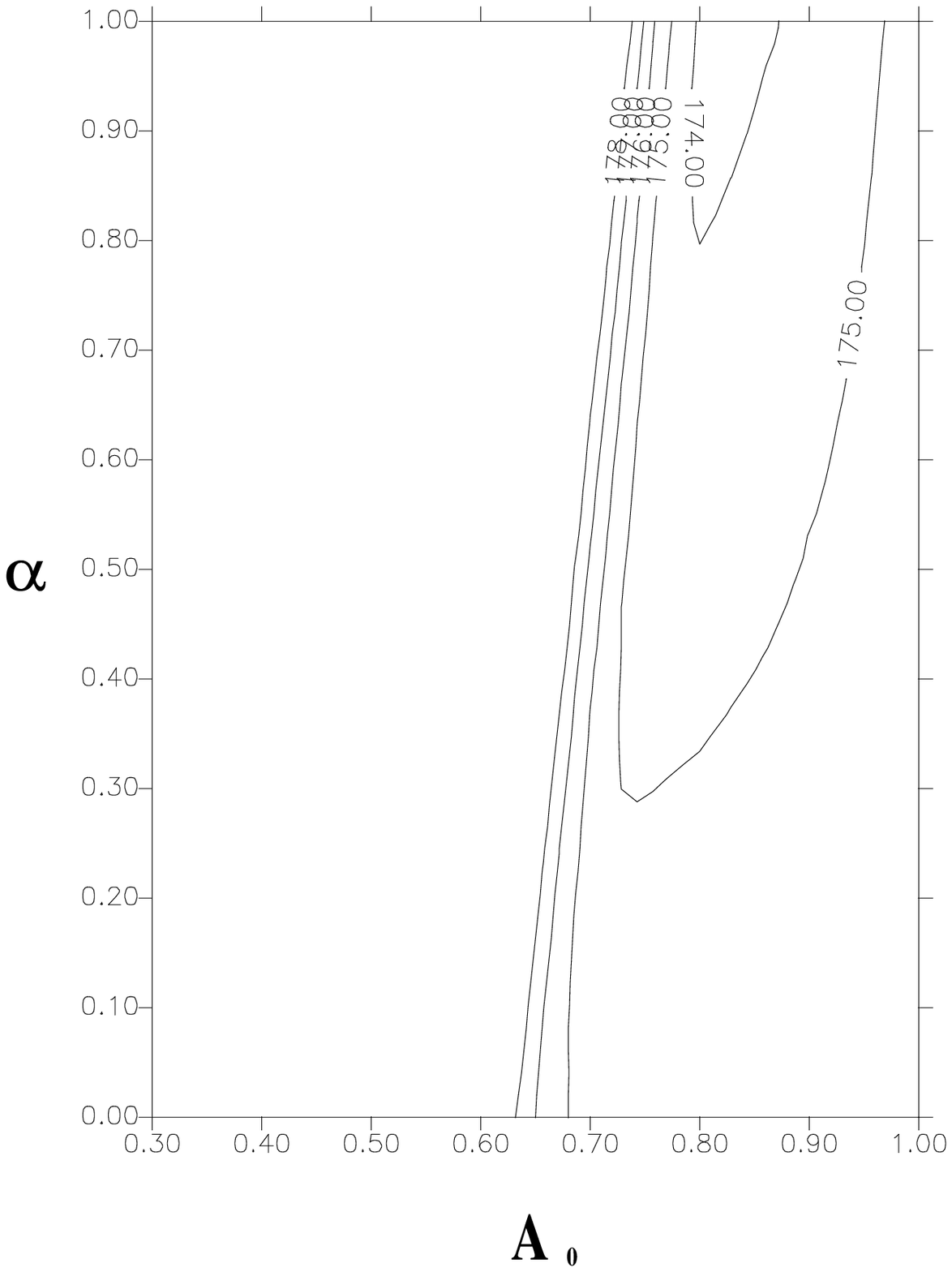}
\caption{Levels of constant $\chi^{2}$ on the plane $(A_0,\alpha)$ for Generalized Chaplygin
Gas model, Gold sample, marginalized over ${\cal M}$, and $\Omega_m$.
The figure shows preferred values of $A_0$ and $\alpha$.}
\label{fig:19}
\end{figure}
 
\begin{figure}
\includegraphics[width=0.8\textwidth]{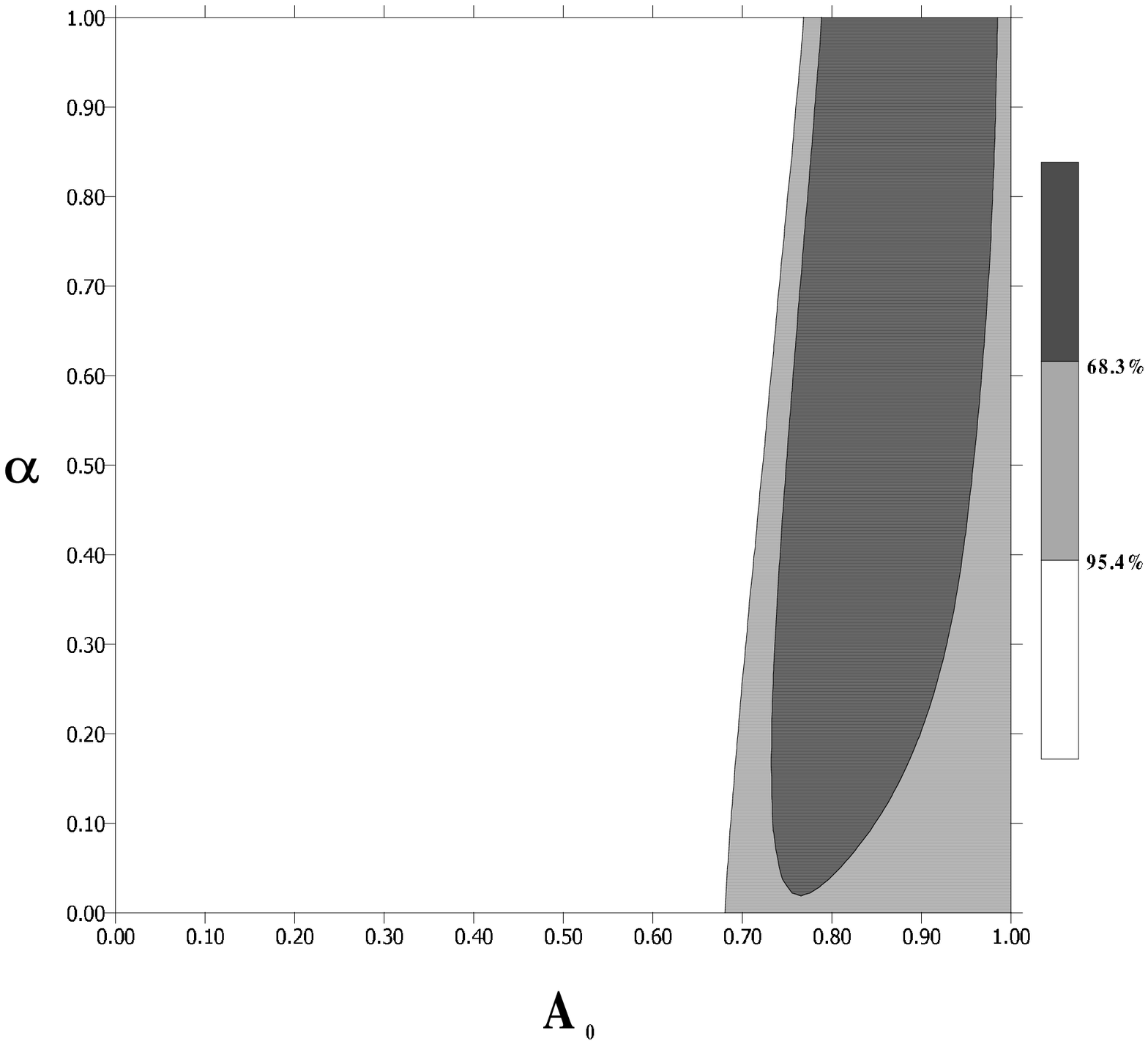}
\caption{Confidence  levels on the plane $(A_0,\alpha)$ for Generalized Chaplygin Gas model,
Gold sample, marginalized over ${\cal M}$, and $\Omega_m$. The figure shows the
ellipses of preferred values of $A_0$ and $\alpha$.}
\label{fig:20}
\end{figure}

\section{Fits to $A_0$ and $\alpha$ parameters}
 
\subsection{Samples used}

Supernovae surveys (published data) have already five years long history.
Beginning with first published samples other data sets have been
produced either by correcting original samples for systematics or by
supplementing them with new supernovae (or both).
It is not our intention here to suggest a distinguished role to any one
of these data sets. Therefore, in our analysis we decided to use a collection
of samples from all existing supernovae data.

Latest  data was compiled by Riess et al. (2004).
However, for comparision and ilustration we analyse three other main samples
of supernovae. Such investigation seems to be useful because it is pointed out
in the literature that result of the analysis with different SNIa sample give
often different results (see for example God{\l}owski Szyd{\l}owski \& Krawiec
2003, Choudhury \& Padmanabhan 2004).
 
Samples from the original Perlmutter et al. (1999) data chosen for the analysis 
comprise the full sample reported by Perlmutter (sample A) and a sub-sample 
after excluding two outliers differing the most from the average lightcurve 
and two outliers claimed to be likely reddened (sample C). Although the 
outliers often suggest statistical inhomogeneity of the data (and some hints 
suggesting the necessity of removing them from sample A exist) there is always 
a danger that removal of outliers is to some extent subjective. Therefore we 
retained the full sample A in our analysis. The Perlmutter data were gathered 
some years ago, hence one should also refer to more recent supernovae data as 
well.
 
Recently Knop et al. (2003) have reexamined the Permutter's data with 
host-galaxy extinction correctly assessed. From the Perlmutter's sample they
chose only these supernovae which were spectroscopically safely identified
as type Ia and had reasonable color measurements. They also included eleven new
high redshift supernovae and a well known sample with low redshift supernovae.
In Knop et al. (2003) a few subsamples have been distinguished. We considered
two of them. The first is a subset of 58 supernovae with corrected extinction
(Knop subsample 6; hereafter K6) and the second is that of 54 low extinction
supernovae (Knop subsample 3; hereafter K3). Samples C and K3 are similarly
constructed  as containing only low extinction supernovae.
The advantage of the Knop sample is that Knop's discussion of extinction
correction was very careful and
as a result his sample has extinction  correctly applied.

Another sample was recently presented by Tonry et al. (2003) who collected a 
large number of supernovae data published by different authors and added eight 
new high redshift SN Ia. This sample of 230 SN Ia was re-calibrated with a 
consistent zero point. Wherever possible the extinction estimates and distance 
fitting were recalculated. Unfortunately, one was not able to do so for the 
full sample (for details see Table~8 in Tonry et al. (2003)). This sample was 
further improved by Barris et al. (2003) who added 23 high redshift supernovae 
including 15 at $z \ge 0.7$ thus doubling the published record of objects at 
these redshifts. Despite of the above mentioned problems, the analysis of our 
model using this sample of supernovae could be interesting. Hence for 
comparison, we decided to repeat our analysis with the Tonry/Barris sample.
 
We decided to analyze two Tonry/Bariss subsamples. First, we considered the 
full Tonry/Barris sample of 253 SNe Ia (hereafter sample TBI). The sample 
contains 218 SNe Ia with low extinction. Because Tonry's sample has a lot of 
outliers especially at low redshifts, we decided to analyze the sample where 
all low redshift ($z<0.01$) supernovae were excluded. In the sample of 193 
supernovae all SN Ia with low redshift and high extinction were removed
(hereafter sample TBII).
 
Tonry et al. (2003) and Barris et al. (2003) presented the data of redshifts 
and luminosity distances for their supernovae sample. Therefore, 
Eqs. (\ref{m-z}) and (\ref{intercept}) should be modified appropriately 
(Williams et al. 2003):
\begin{equation}
\label{eq:13a}
m-M = 5\log_{10}(D_L)_{\mathrm{Tonry}}-5\log_{10}65 + 25
\end{equation}
and
\begin{equation}
\label{eq:13b}
{\cal M}=-5\log_{10}H_0+25.
\end{equation}
For the Hubble constant $H_0=65$ km s$^{-1}$ Mpc$^{-1}$ one gets  ${\cal M}=15.935$.

Recently Riess et al. (2004) significantly improved the former Riess sample.
They discovered 16 new type Ia Supernovae. It should be noted that 6 of these 
objects have $z>1.25$ (out of total number of 7 object with so high redshifts).
Moreover, they compiled a set of previously observed SNIa relying on large, 
published samples, whenever possible, to reduce systematic errors from 
differences in calibrations. With this enriched sample it became possible to 
test our prediction that distant supernovae in GCG cosmology
should be brighter than in  $\Lambda$CDM model (see discussion below).
This is the reason why we repeated our analysis with new Riess sample.
 
The full Riess sample contains 186 SNIa (``Silver'' sample). On the base of 
quality of the spectroscopic and photometric record for individual Supernovae, 
they also selected more restricted ``Gold'' sample of 157 Supernovae. 
We have separately analyzed $\Lambda$CDM model  for supernovae  with $z<1$ 
and for all SNIa  belonging to the Gold sample.
 
\subsection{Cosmological models tested}
 
On these samples we have tested Generalized Chaplygin gas cosmology in three
different classes of models with (1) $\Omega_m= 0.3$, $\Omega_{Ch}= 0.7$;
(2) $\Omega_m= 0.05$, $\Omega_{Ch}= 0.95$ and (3) $\Omega_m = 0$, $\Omega_{Ch} =  1$.
We started with a fixed value of ${\cal M} =  -3.39$ modifying this assumption accordingly
while analyzing different samples.
 
The first class was chosen as representative of the standard knowledge of
$\Omega_m$ (baryonic plus dark matter in galactic halos (Peebles \& Ratra 2003))
with Chaplygin gas responsible for the missing part of closure density (the dark energy).
 
In the second class we have incorporated (at the level of $\Omega_m$) the prior knowledge
about baryonic content of the Universe (as inferred from the BBN considerations).
Hence this class is representative of the models in which Chaplygin gas is allowed to
clump and is responsible both for dark matter in halos as well as its diffuse part
(dark energy).
 
The third class is a kind of toy model -- the FRW Universe filled completely with
Chaplygin gas. We have considered it mainly in order to see how sensitive the SNIa
test is with respect to parameters identifying the cosmological model.
 
Finally, we analyzed the data without any prior assumption about $\Omega_m$.
 
\subsection{Results}
 
In the first class of models best fit (with fixed value of ${\cal M} =  -3.39$) from the
sample A is $(\alpha =  1, A_0 =  0.96)$ at the $\chi^2 = 95.8$.
Sample C gives the best fit of $(\alpha =  0.95, A_0 =  0.95)$ at the $\chi^2 =  53.6$.
 
In the second class Sample A gives the best fit of $(\alpha =  1, A_0 =  0.80)$
at the $\chi^2 =  95.4$ whereas the sample C gives the best fit
$(\alpha =  0.51, A_0 =  0.73)$ at the $\chi^2 =  53.7$.
 
Finally, in the third class the sample
A again gives the best fit of $(\alpha =  1, A_0 =  0.77)$ at the $\chi^2 =  95.4$ whereas
the sample C gives the best fit $(\alpha =  0.42, A_0 =  0.69)$ at
the $\chi^2 =  53.7$.
 
It should be noted, however that the fitting procedure
for sample C prefers
${\cal M} =  -3.44$ instead of ${\cal M} =  -3.39$ as for sample A. If one takes this
value the results for sample C will change respectively and then for the first class
$A_0 =  1$ (at $\chi^2 =  53.5$) what means (see equation (\ref{Hubble}))
that $\alpha$ can be arbitrary and the problem is effectively equivalent to
the model with  cosmological constant. Analogously, for the second class
$A_0 =  0.83$ $\alpha= 1$ (at $\chi^2 =  52.9$), while for third class
$A_0 =  0.80$ $\alpha= 1$ (at $\chi^2 =  52.9$). This indicates clearly that model
parameters, especially $\alpha$ strongly depend on the choice of ${\cal M}$.
 
Separately we analyzed the data without any prior assumption about $\Omega_m$.
For the sample A we obtain as a best fit
(minimizing $\chi^2$) $\Omega_m =  0. $, $(\alpha =  1, A_0 =  0.77)$
at the $\chi^2 = 95.4$. For sample C assuming ${\cal M}= -3.39$
we obtain $\Omega_m =  0.27 $, $(\alpha =  1, A_0 =  0.93)$
with $\chi^2 = 53.6$ while for ${\cal M}= -3.44$ the best fit
gives $\Omega_m =  0. $, $(\alpha =  1, A_0 =  0.80)$ with $\chi^2 = 52.9$.
 
Using minimization procedure, based on likelihood function,
joint marginalization over $A_0$ and $\alpha$ gives the following results.
For the sample A we obtain $\Omega_{Ch} =  0.82$ (hence $\Omega_m =  0.18 $),
with the limit $\Omega_{Ch} \geq 0.76$ at the confidence level $68.3 \%$
and $\Omega_{Ch} \geq 0.69 $ at the confidence level $95.4 \%$.
For the sample C we obtain (for ${\cal M} = -3.39$) $\Omega_{Ch} =  0.76$
(hence $\Omega_m =  0.24 $), with the limit $\Omega_{Ch} \epsilon (0.69,0.94)$ on the
confidence level $68.3 \%$ and $\Omega_{Ch} \geq 0.62 $ at the confidence level $95.4 \%$.
For ${\cal M} = -3.44$ we obtain: $\Omega_{Ch} =  0.84$ (hence $\Omega_m =  0.16 $),
with the limit $\Omega_{Ch} \epsilon (0.79,0.98)$ on the confidence level $68.3 \%$ and
$\Omega_{Ch} \geq 0.69 $ at the confidence level $95.4 \%$.
 
One could see that results are different for different values of the intercept
${\cal M}$ in each sample. Therefore we additionally analyzed our samples
marginalized over ${\cal M}$. The results are displayed in Table 1.
First rows for each sample correspond to no prior on $\Omega_m$ assumed.
Two fitting procedures were used: $\chi^2$ fitting (denoted as BF) and maximum
l ikelihood method (denoted L).
 
Table 2 contains the results of joint marginalization over $\Omega_m$,
$A_0$ and $\alpha$.
For sample $A$ we obtain $\Omega_{Ch} =  0.83$ (hence $\Omega_m =  0.17 $), 
while for sample $C$ we obtain $\Omega_{Ch} =  0.85$
(hence $\Omega_m =  0.15 $). For $A_0$ and $\alpha$ we obtain values
$(\alpha =  0, A_0 = 0.83)$ for sample A, while $(\alpha =  0, A_0 = 0.86)$
for sample C. With the marginalization procedure we can also obtain one 
dimensional probability distribution function (PDF) for $A_0$ and $\alpha$ 
for particular class models with fixed $\Omega_m$ (Table 3).

From the above mentioned analysis we concluded that
$\Omega_m$ and $A_0$ parameters derived from samples $A$ and $C$ are similar.
For $\Omega_m$ fixed $A_0$ increases
with increasing $\Omega_m$. The estimates of
$\alpha$ parameter are different for each of two above mentioned samples,
but unfortunately errors are big.
Discrepancy between the best fitting procedure
and minimization procedure (based on likelihood function) increases
with the number of parameters fitted. The minimization procedure seems to be
more appropriate in the context of our problem.
 
One can see from the Table 1 that using the Knop's samples had not influenced
conclusions in a significant way. However, the errors of parameter estimation decreased
noticeably (see Tables 2 and 3). Minimization procedure prefers (especialy
for sample K3) $\alpha$ close to zero. The exception is the model with $\Omega_m=0$
where $\alpha=0.3$ and $\alpha=0.71$ are obtained for the samples K3 and K6 respectively.

The above mentioned results for the Knop sample K3 are illustrated on figures Fig. 1 - 17.
In Fig.1 we present residual plots of redshift-magnitude relations
between the Einstein-de Sitter model (represented by zero line) the best-fitted Generalized
Chaplygin Gas model  with $\Omega_m= 0.3, \Omega_{Ch}= 0.7)$ (middle curve) and the flat
$\Lambda$CDM model with $\Lambda=0.75$ and $\Omega_{\mathrm{m}}=0.25$
(Knop et al. 2003)) -- upper curve. One can observe that systematic deviation between
$\Lambda$CDM model and Generalized Chaplygin Gas  model gets larger at higher redshifts.
The Generalized Chaplygin Gas model predict that high redshift supernovae
should be brighter than that predicted with $\Lambda$CDM model.
 
Levels of constant $\chi^{2}$ on the $(A_0,\alpha)$ plane for Generalized
Chaplygin Gas model  with $\Omega_m= 0.3, \Omega_{Ch}= 0.7)$, marginalized
over ${\cal M}$ are presented in Figure 2. The figure shows preferred values
of $A_0$ and $\alpha$. Figure 3 displays confidence levels on the $(A_0,\alpha)$
plane for Generalized Chaplygin Gas model with
$\Omega_m= 0.3, \Omega_{Ch}= 0.7)$ , marginalized over ${\cal M}$.
This figure shows the ellipses of preferred values of $(A_0$ and $\alpha)$.
Similar results for the models with $\Omega_m= 0.05, \Omega_{Ch}= 0.95$ and
$\Omega_m= 0, \Omega_{Ch}= 1$ are presented in the figures 4 - 9 respectively.
 
Separately we repeated our analysis without prior assumptions on $\Omega_m$. The density
distribution (one dimensional PDF) for model parameters obtained by marginalization over
remaining parameters of the model are presented in Figures 10-12.
 
 Residuals (in mag) between the Einstein-de Sitter model (zero line), flat $\Lambda$CDM
model -- upper curve -- and the best-fitted Generalized Chaplygin Gas model
(without prior assumptions on $\Omega_m$) -- middle curve -- are presented on Figure 13,
while levels of constant $\chi^{2}$ and confidence levels on the $(A_0,\alpha)$ plane
(marginalized over ${\cal M}$) are presented on Figs. 14 and 15.
 
One should notice that as a best fit we obtain
$\Omega_m= 0, \Omega_{Ch}= 1, A_0=0.85, \alpha=1$ i.e. results are the same as for a toy
model with Chaplygin gas only ($\Omega_{Ch}= 1$). Formally, we could have analyzed models
with $\alpha>1$. However, due to large error in estimation of the $\alpha$ parameter,
it does not seem reasonable to analyze such a possibility with current
supernovae data.
 
Levels of constant $\chi^{2}$ on the $(\Omega_m,\alpha)$ plane for Generalized Chaplygin Gas
model, marginalized over ${\cal M}$ and $A_0$  are presented on Figure 16, while confidence
levels on the $(\Omega_m, \alpha)$ plane  (marginalized over ${\cal M}$, and $A_0$) are
presented on Figure 17. This figure shows that all three model with $\Omega_{Ch}= 1$,
$\Omega_{Ch}=0.95$, and $\Omega_{Ch}=0.7$, are statistically admissible by
current Supernovae data.
 
The results of similar analysis obtained with Tonry/Barris sample are
similar to those obtained with previous samples.
For example TBII sample gives the best fit: $\Omega_m= 0, \Omega_{Ch}= 1,
A_0=0.78, \alpha=1$ i.e. nearly the same as in the case of K3 sample.
 
Joint marginalization over parameters gives the following results:
$\Omega_{Ch} =  1.00$ (hence $\Omega_m =  0.0 $), with the limit
$\Omega_{Ch} \geq 0.79$ at the confidence level of $68.3 \%$ and
$\Omega_{Ch} \geq 0.67 $ at the confidence level of $95.4 \%$.
$(\alpha =  1.0, A_0 =  0.81)$ with the limit $\alpha \in (0.40, 1.)$ and
$A_0 \in (0.74, 0.93)$ at the confidence level of $68.3 \%$ and
$\alpha \in (0.06, 1.)$ and $A_0 \in (0.70,1.00)$ at the confidence level of
$95.4 \%$.
 
However with minimization procedure we find important difference beetween
results obtained with Tonry/Barris sample and those
obtained with Perlmutter and Knop samples. Minimization procedure (only
except the model with fixed $\Omega_m=0.3$) performed on Tonry/Barris data
gives $\alpha=1$. It is significantly different from the result obtained for
$C$ Perlmutter's and Knop's samples where minimization procedure preffered small
values of $\alpha$ parameter. Also the Tonry /Barris sample preffered value of
$\Omega_m=0$ while Perlmutter and Knop samples sugested $\Omega_m$ is close to
zero, what indicates that barionic component is small and in agreement with BBN.
 
The new Riess sample leads to the results which are similar to these obtained with
Tonry/Bariss sample. However the errors in estimation of the parameters are lower.
For the Gold sample, joint marginalization  over parameters gives the following
results: $\Omega_{Ch} =  1.00$ (hence $\Omega_m =  0.0 $), with the limit
$\Omega_{Ch} \geq 0.80$ at the confidence level of $68.3 \%$ and $\Omega_{Ch} \geq 0.69$
at the confidence level of $95.4 \%$. $(\alpha =  1.0, A_0 =  0.83)$ with the
limit $\alpha \in (0.36, 1.)$ and $A_0 \in (0.76, 0.94)$ at the
confidence level of $68.3 \%$ and $\alpha \in (0.05, 1.)$ and
$A_0 \in (0.72,1.00)$ at the confidence level of $95.4 \%$.
 
As one can see from the Figure 18 the differences between the results obtained
in both cases are small (however the result obtained with the full sample
leads to the prediction of brighter distant supernovae than in the case with $z<1$ SNIa.)
Residuals (in mag) between the Einstein-de Sitter model (zero line),
flat $\Lambda$CDM model (two upper curves: for SNIA with $z<1$ - higher curve - and for
all supernovae Ia belonging to the sample - lower curve) and the best-fitted Generalized
Chaplygin Gas model (without prior assumptions on $\Omega_m$) (middle curve)
are presented on Figure 18. Figure 18  shows  that most distant supernovae are
actually brighter than predicted in  $\Lambda$CDM model.
This is in agreement with prediction of the Generalized Chaplygin Gas cosmology.
 
The levels of constant $\chi^{2}$ and confidence levels on the $(A_0,\alpha)$
plane (marginalized over ${\cal M}$) are presented on Figs. 19 and 20.
Fig.20 shows that confidence levels on the $(A_0,\alpha)$ (for Gold Sample)
plane are comparable at the the $95.4 \%$ confidence level with the results
obtained on the Knop's sample. However the preferred values of
$\alpha$ are different.

\subsection{Flat prior relaxed}
 
We extended our analysis  by adding a curvature term
to the original GCG model. Then in equation (\ref{LD}) we must take into
account the $\Omega_{k}$ term. For statistical analysis we restricted the
values of the   $\Omega_{m}$ parameter to the interval $[0,1]$, $\Omega_{Ch}$
to the interval $[0,2]$ and $\Omega_{k}$ was obtained from the constraint
 $\Omega_{m}+\Omega_{ch}+\Omega_{k}=1$. However, the cases $\Omega_{k} <-1$
were excluded from the analysis. The results are presented in Table IV and V.
 
In the model without prior assumptions on $\Omega_m$ we obtain with Knop's sample
$\Omega_{k}=-0.19$ as a best fit, while with maximum likelihood method
prefers $\Omega_{k}=-0.60$. However, the model with priors
on $\Omega_m$ or $\Omega_{Ch}$ the maximum likelihood method prefers the
universe much ``closer'' to the flat one. Specifically, for the "toy" model
with Chaplygin Gas only one gets $\Omega_{k}=0.10$ and $\Omega_{k}=0.05$
for the model with baryonic content only i.e. $\Omega_{m}=0.05$.
One should emphasize that even though we allowed $\Omega_{k} \ne 0$
the preferred model of the universe is nearly a flat one, which is in
agreement with CMBR data. It is an advantage of our GCG model as compared with
$\Lambda$CDM model where in Perlmutter et al. (1999) high negative value of
$\Omega_{k}$ was obtained as a best fit, although zero value of $\Omega_{k}$
was statistically admissible. In order to find the curvature of the Universe
they additionally used the data from CMBR and extragalactic astronomy.

Density distribution functions (one dimensional PDF) for model parameters
obtained by marginalization over remaining parameters of the model
are presented in Figures 21-25. For $\Omega_{k}$ we obtain the limit
$\Omega_{k} \in (-0.98,-0.22)$  at the confidence level $68.3 \%$
and  $\Omega_{k} \in (-1,0.23)$  at the confidence level $95.4 \%$.
For  $\alpha$ and  $A_0$ parameters we obtain the following results:
$\alpha =  0.$ and $A_0 = 0.89$ with the limit $\alpha \in (0,0.64)$
and $A_0 \in (0.82,1)$ at the confidence level $68.3 \%$ and
$\alpha \in (0., 0.95)$ and $A_0 \in (0.73,1.)$ at the
confidence level $95.4 \%$. For the density parameter $\Omega_{Ch}$
we obtain the limit  $\Omega_{Ch} \in (0.87,1.51)$ at the confidence
level $68.3 \%$ and  $\Omega_{Ch} \in (0.61,1.79)$ at the confidence
level $95.4 \%$. For the density parameter $\Omega_{m}$ we obtain the
limit  $\Omega_{m} <0.29$  at the confidence level $68.3 \%$ and
$\Omega_{m} <0.53$ at the confidence level $95.4 \%$.
 
Our main result is that the preference of the nearly flat universe is confirmed
with the new Riess sample. In the model without a prior
assumption on $\Omega_m$ we obtain $\Omega_{k}=-0.12$ as a best fit,  with the
Gold sample  while maximum likelihood method prefers $\Omega_{k}=-0.32$ i.e
the Gold sample gives even "more flat" universe than Knop's sample.
The models with priors on $\Omega_m$  give also very similar results
when we analyse Knop and Riess samples.
One can see that estimation of other models parameters give similar
result for both samples only with exception for parameter $\alpha$.
Specifically, for the ``toy"  and ``baryonic" models
the maximum likelihood method prefers the universe with non zero
parameter $\alpha$ like for "flat universe".
One can see that, with flat prior relaxed, when we analyse the Gold sample,
the errors in estimation of the model parameter significantly decrase with
comparision to the case of the Knop sample.

\section{Generalized Chaplygin Gas model in perspective of SNAP data}
 
In the near future the SNAP mission is expected to observe about 2000 SN
Ia supernovae each year, over a period of three years
\footnote{http://www-supernova.lbl.gov, http://snfactory.lbl.gov}.
Therefore it could be possible to discriminate between various cosmological
models since errors in the estimation of model parameters would decrease
significantly. We tested how a large number of new data would influence the
errors in estimation of model parameters. We assumed that the Universe is
flat and tested three classes of cosmological models. In the first
$\Lambda$CDM model we assumed that $\Omega_{\mathrm{m}}=0.25$, and
$\Omega_{\Lambda}=0.75$ and ${\cal M}=-3.39$ (Knop et al. 2003).
Second class was representative of the so called Cardassian models
(Freese \& Lewis 2002) with parameters $\Omega_{m}=0.42$, $\Omega_{card}=0.52$
and $n=-0.77$ as obtained in (God{\l}owski, Szyd{\l}owski \& Krawiec 2004).
Let us note, that technically i.e. at the level of tests like the Hubble
diagram Cardassian models are equivalent to quintessence models. The
difference is in the underlying philosophy: quintessence assumes exotic
dark energy component with hydrodynamical equation of state in ordinary FRW
model while the Cardassian Universe assumes modification of the Friedman
equation (which can be either due to exotic matter component or due to
modification of gravity law). The last model is Generalized Chaplygin Gas
Model with parameters obtained in the present paper as  best fits for the K3
sample ($\Omega_m=0$, $A_0=0.85$, $\alpha=1.$). These values are in agreement
with results of the analysis performed on Tonry/Barris and Riess samples.
Alternatively, we also test the  Generalized Chaplygin Gas Model with
small value of $\alpha$ parameter suggested by analysis of the
Perlmmutter and Knop samples  ($\Omega_m=0$, $A_0=0.76$, $\alpha=0.40$).
For three above mentioned models we generated a sample of 1915 supernovae
(Samples X1,X2,X3a,X3b respectively) in the redshift range
$z \epsilon [0.01,1.7]$ distributed according to predicted SNAP data
(see Tab I of Alam et al. (2003)). We assumed Gaussian distribution of
uncertainties in the measurement of m and z. The errors in redshifts $z$
are of order $1\sigma=0.002$ while uncertainty in the measurement of
magnitude $m$ is assumed $1\sigma=0.15$. The systematic uncertainty is
$\sigma_{sys}=0.02$ mag at $z=1.5$ (Alam et al. 2003). Hence one can assume
that $\sigma_{sys}(z)=(0.02/1.5)z$ in first approximation. For such generated
sample we repeated our analysis. The result of our analysis is presented on
figures Fig.26-29. On these figures we present confidence  levels on the
plane $(A_0,\alpha)$ for sample of simulated SNAP data. The figures show the
ellipses of preferred values of $A_0$ and $\alpha$. It is easy to see that
with the forthcoming SNAP data it will be possible to discriminate between
predictions of $\Lambda$CDM and GCG models. With Cardasian Model the
situation is not so clear, however (see Fig 27 and 28). Note that if
$\alpha \simeq 0.4$ as suggested by analysis of Perlmutter sample C,
(see also: Makler, de Oliveira \& Waga 2003, Avelino et al. 2003, Fabris,
Gonn{\c c}alves \& de Souza 2002, Collistete et al. 2003) then it will be
possible to discriminate between model with Chaplygin gas and Cardassian
model (see Fig 27 and 29). Moreover, it is clear that with the future SNAP
data it would be possible to differentiate between models with various value
of $\alpha$ parameter. This is especially valuable since all analyses
performed so far have had weak sensitivity with respect to $\alpha$.

\section{Conclusions}

It is apparent that Generalized
Chaplygin Gas models have brighter supernovae at redshifts $z>1$. Indeed one
can see on respective figures (Figs 1, 4, 7, 13) that systematic
deviation from the baseline Einstein de Sitter model gets larger at higher
redshifts. This prediction seems to be independent of analysed sample.
 
We obtained that the estimated value of $A_0$ is close to $0.8$ in all 
considered models with exepction the model class (1) ($\Omega_m=0.3$) when
$A_0>0.95$.
Extending our analysis by relaxing the flat prior lead to the result
that even though the best fitted values of $\Omega_k$ are formally non-zero,
yet they are close to the flat case. It should be viewed as an advantage of
the GCG model since in similar analysis of $\Lambda$CDM model in Perlmutter
et al. (1999) high negative value of $\Omega_{k}$ were found to be best
fitted to the data and independent inspiration from CMBR and extragalactic
astronomy has been invoked to fix the curvature problem. Another advantage
of GCG model is that in a natural way we obtained the conclusion that matter
(baryonic) component should be small what is in agreement with prediction
from BBN (Big Bang Nucleosynthesis). Both estimations of $A_0$, $\Omega_k$ and
$\Omega_m$ are independent of the sample used in our analysis.

Our results suggest that SNIa data support the Chaplygin gas (i.e. 
$\alpha = 1$) scenario when the $\chi^2$ best fitting procedure is used.
The minimization procedure performed on Tonry/Barris  and Riess data gives
also $\alpha=1$ (only except the model with fixed $\Omega_m=0.3$).
However, the maximum likelihood fitting with Knop et al.'s sample prefers,
quite unexpectedly, a small value of $\alpha$ or even $\alpha = 0$, i.e.
the $\Lambda$CDM scenario. Please note that small value of $\alpha$
is in agreement with the results obtained from CMBR
(de Bernardis et al. 2000 Benoit et al. 2003,
Hinshaw et al. 2003, Bento, Bertolami \& Sen 2002, Amendola et al. 2003)
and with the recent analysis of Zhu (2004) who, using combined data of X-ray 
gas mass fraction of the galaxy cluster, FR IIB radiogalaxies and combined 
sample Perlmutter et al. (1998) and Riess et al. (1998, 2001), sugested that 
$\alpha$ could be even less than 0.
The results are dependent both on the sample chosen and on the prior
knowledge of $\cal M$ in which the Hubble constant and intrinsic luminosity
of SNIa are entangled. Moreover the observed preference of $A_0$ values
close to 1 means that the $\alpha$ dependence becomes insignificant
(see equation (\ref{Hubble})). It is reflected on one dimensional PDFs for
$\alpha$ which turned out to be flat meaning that the power of the present
supernovae data to discriminate between various Generalized Chaplygin Gas
models (differing by $\alpha$) is weak.
 
However, we argue that with future SNAP data it would be possible to
differentiate between models with various value of $\alpha$ parameter.
Residual plots indicate the differences between $\Lambda$CDM and Generalized
Chaplygin Gas cosmologies at high redshifts.
Therefore one can expect that future supernova experiments (e.g. SNAP) having
access to higher redshifts will eventually resolve the issue whether the dark
energy content of the Universe could be described as a Chaplygin gas.
The discriminative power of forthcoming SNAP data has been illustrated on
respective figures (Fig.26-29) obtained from the analysis on simulated SNAP
data.
 
\begin{figure}
\includegraphics[width=0.8\textwidth]{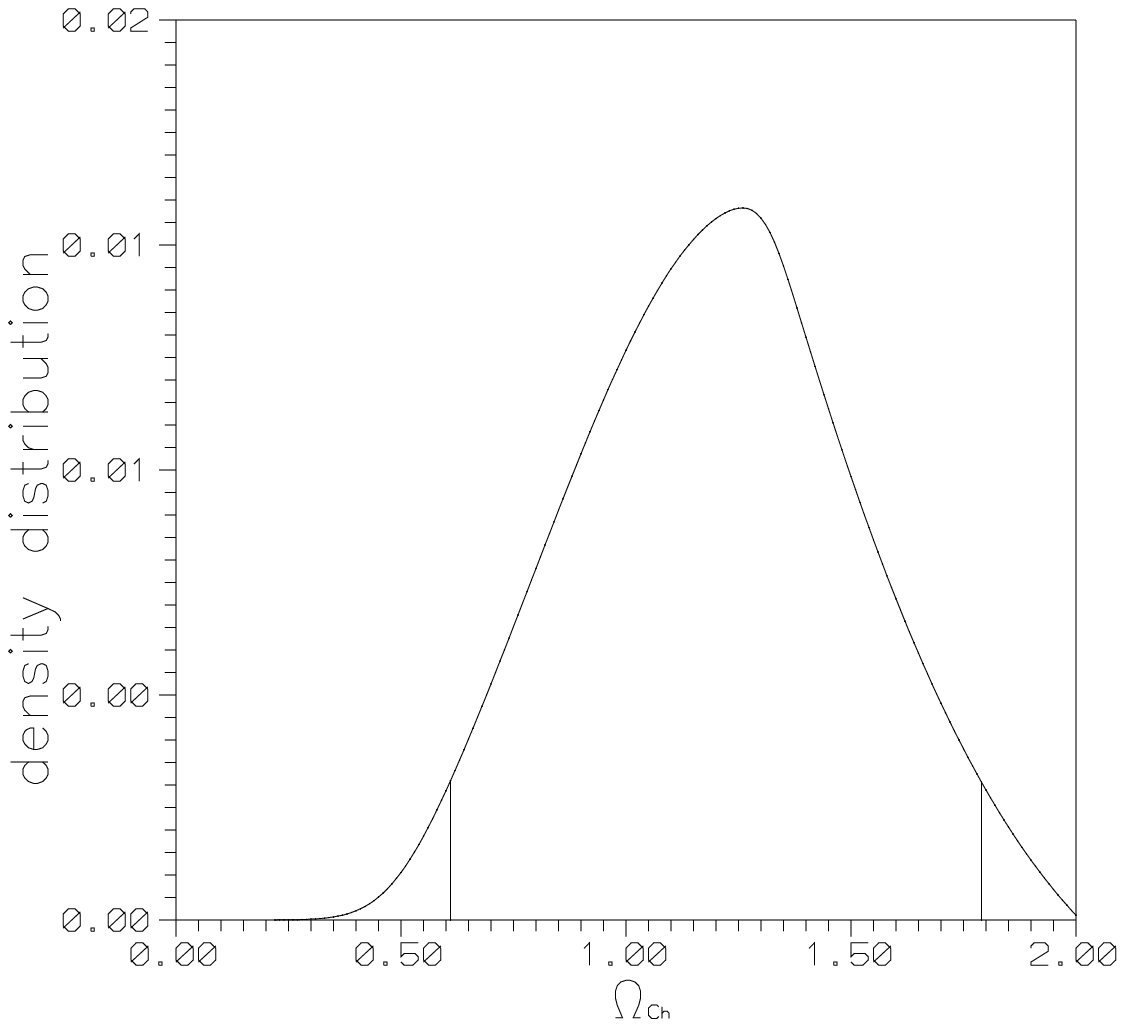}
\caption{The density distribution (one dimensional PDF) for $\Omega_{Ch}$ obtained from
sample K3 by marginalization over remaining parameters of the model. We obtain the limit
$\Omega_{Ch} \epsilon (0.61, 1.79)$ at the confidence level $95.4 \%$.
(Non-flat GCG model.)}
\label{fig:21}
\end{figure}
 
\begin{figure}
\includegraphics[width=0.8\textwidth]{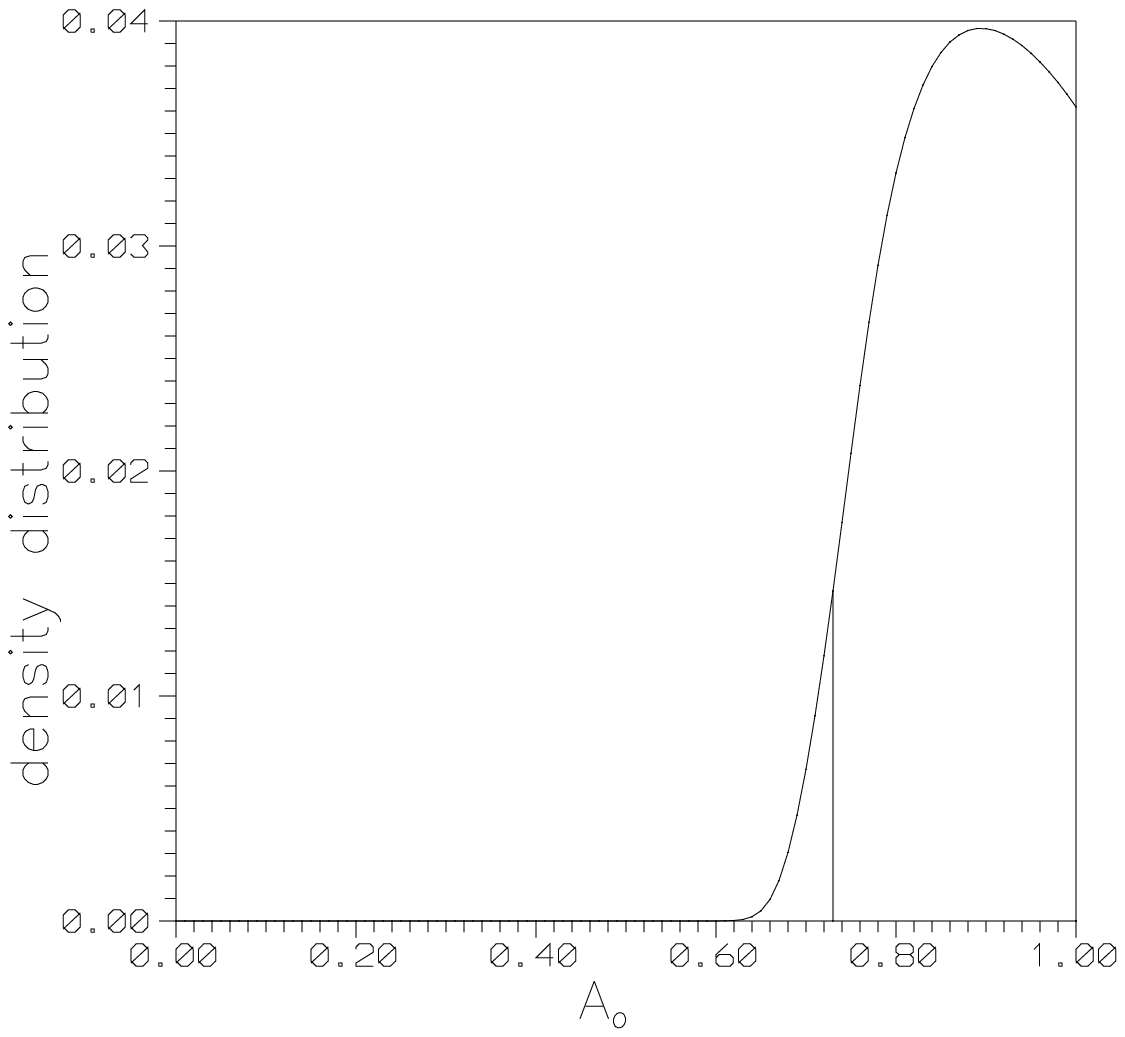}
\caption{The density distribution (one dimensional PDF) for $A_0$ obtained from
sample K3 by marginalization over remaining parameters of the model. We obtain the
limit $A_0 \epsilon (0.73,1)$ at the confidence level $95.4 \%$.(Non-flat GCG model.)}
\label{fig:22}
\end{figure}
 
\begin{figure}
\includegraphics[width=0.8\textwidth]{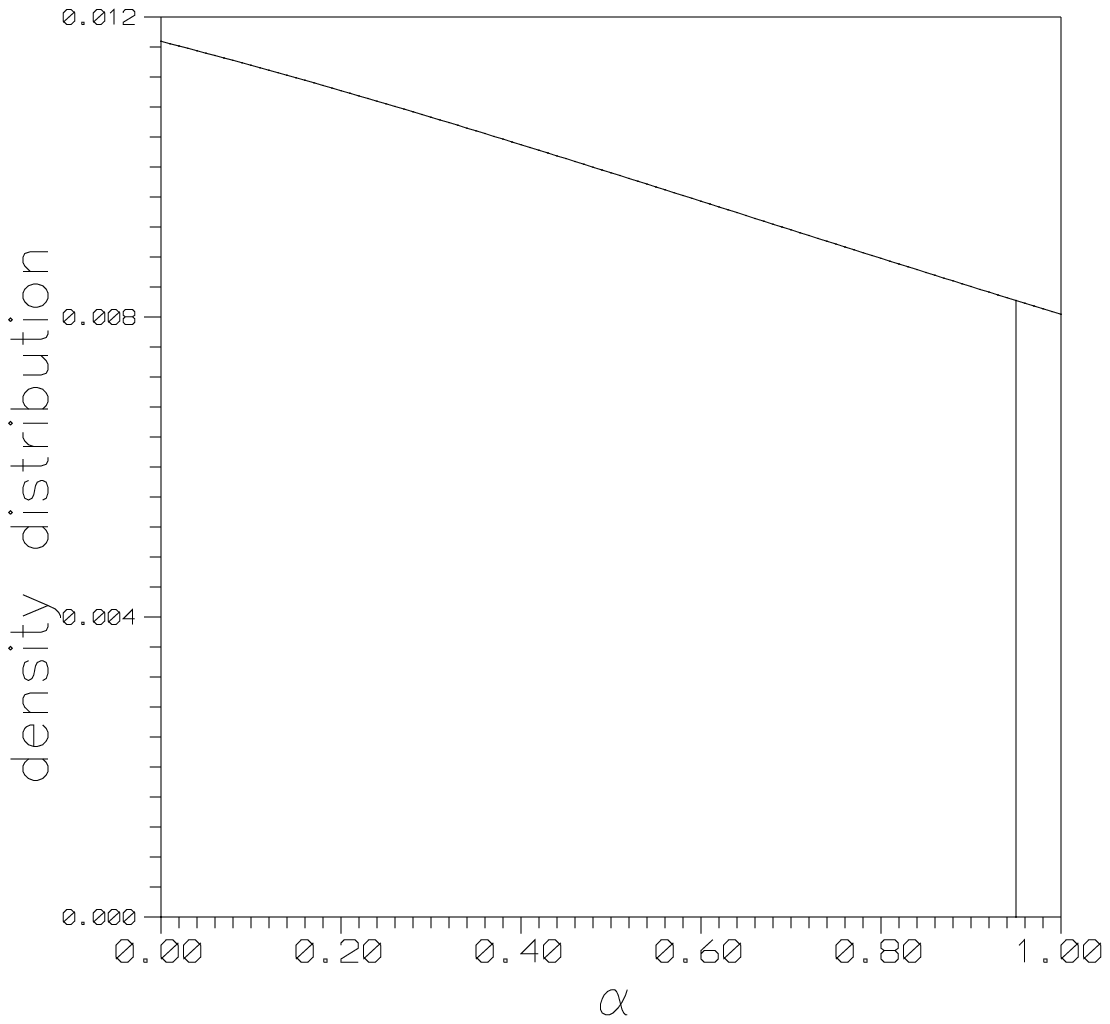}
\caption{The density distribution (one dimensional PDF) for $\alpha$ obtained from
sample K3 by marginalization over remaining parameters of the model.
We obtain the limit $\alpha < 0.95$ at the confidence level $95.4 \%$.(Non-flat GCG model.)}
\label{fig:23}
\end{figure}
 
\begin{figure}
\includegraphics[width=0.8\textwidth]{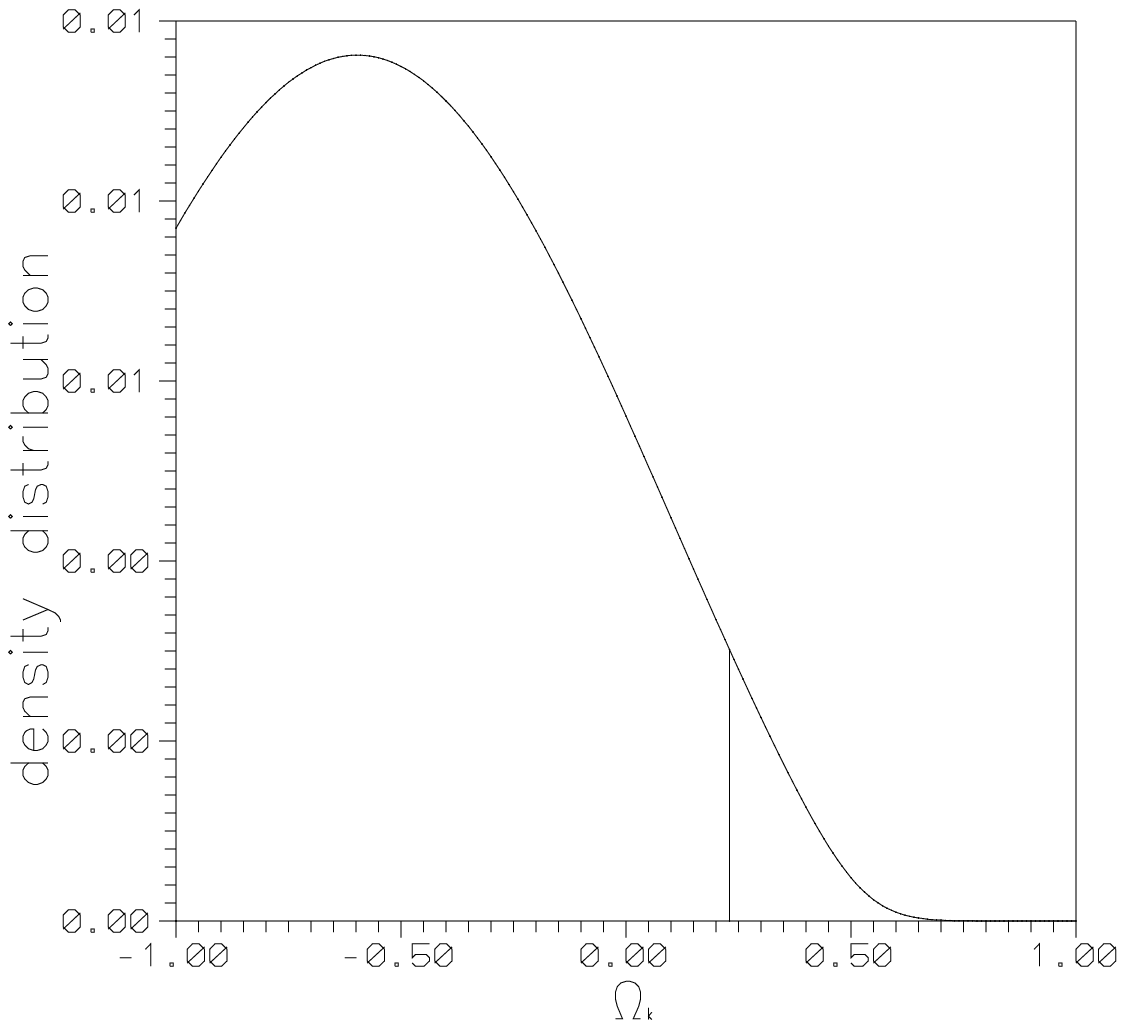}
\caption{The density distribution (one dimensional PDF) for $\Omega_{k}$ obtained from
sample K3 by marginalization over remaining parameters of the model. We obtain the
limit $\Omega_{k} \epsilon (-1,0.23)$ at the confidence level $95.4 \%$.
(Non-flat GCG model.)}
\label{fig:24}
\end{figure}
 
\begin{figure}
\includegraphics[width=0.8\textwidth]{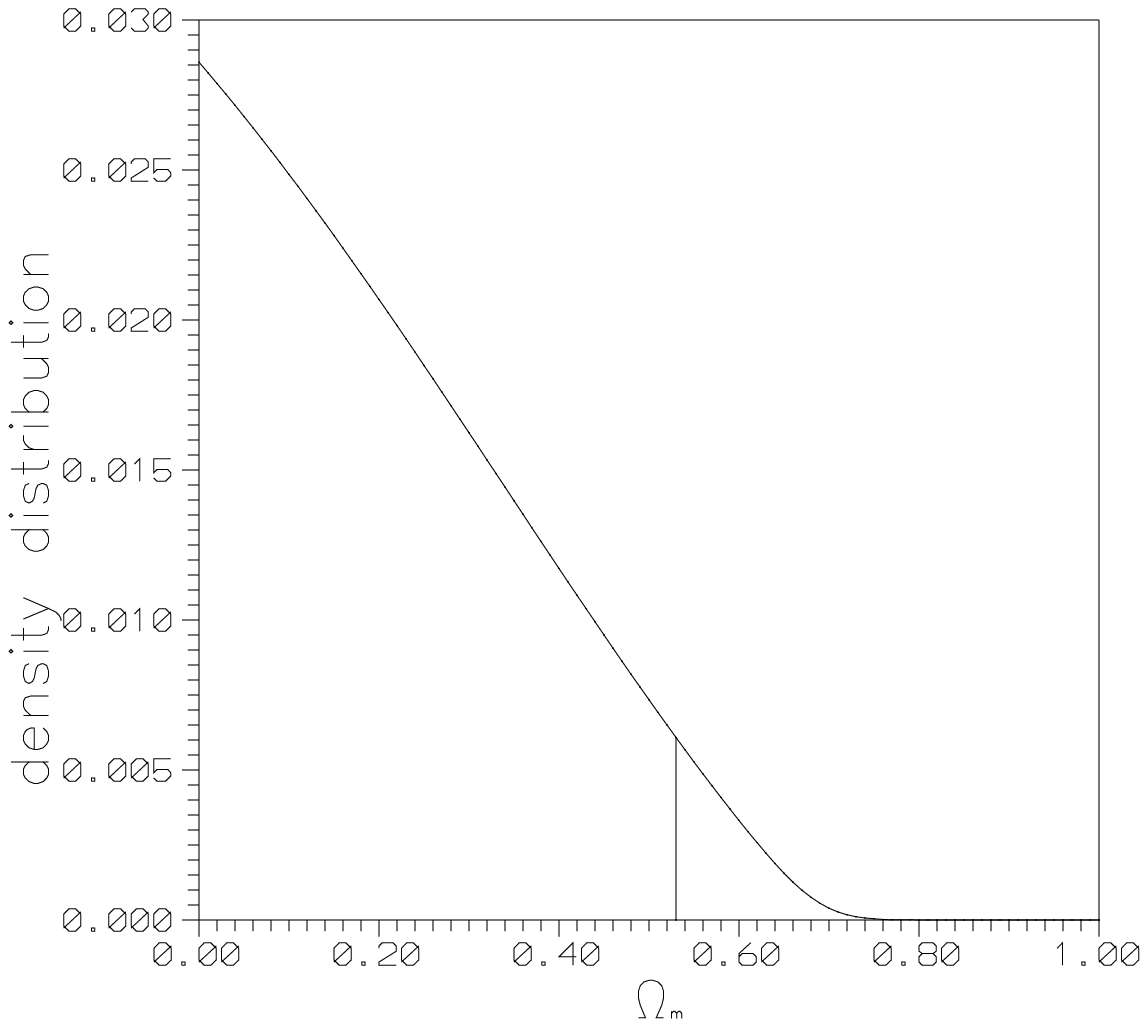}
\caption{The density distribution (one dimensional PDF) for $\Omega_{m}$ obtained from
sample K3 by marginalization over remaining parameters of the model.
We obtain the limit $\Omega_{m} \epsilon( 0,0.53)$ at the confidence level $95.4 \%$.
(Non-flat GCG model.)}
\label{fig:25}
\end{figure}
 
\begin{figure}
\includegraphics[width=0.8\textwidth]{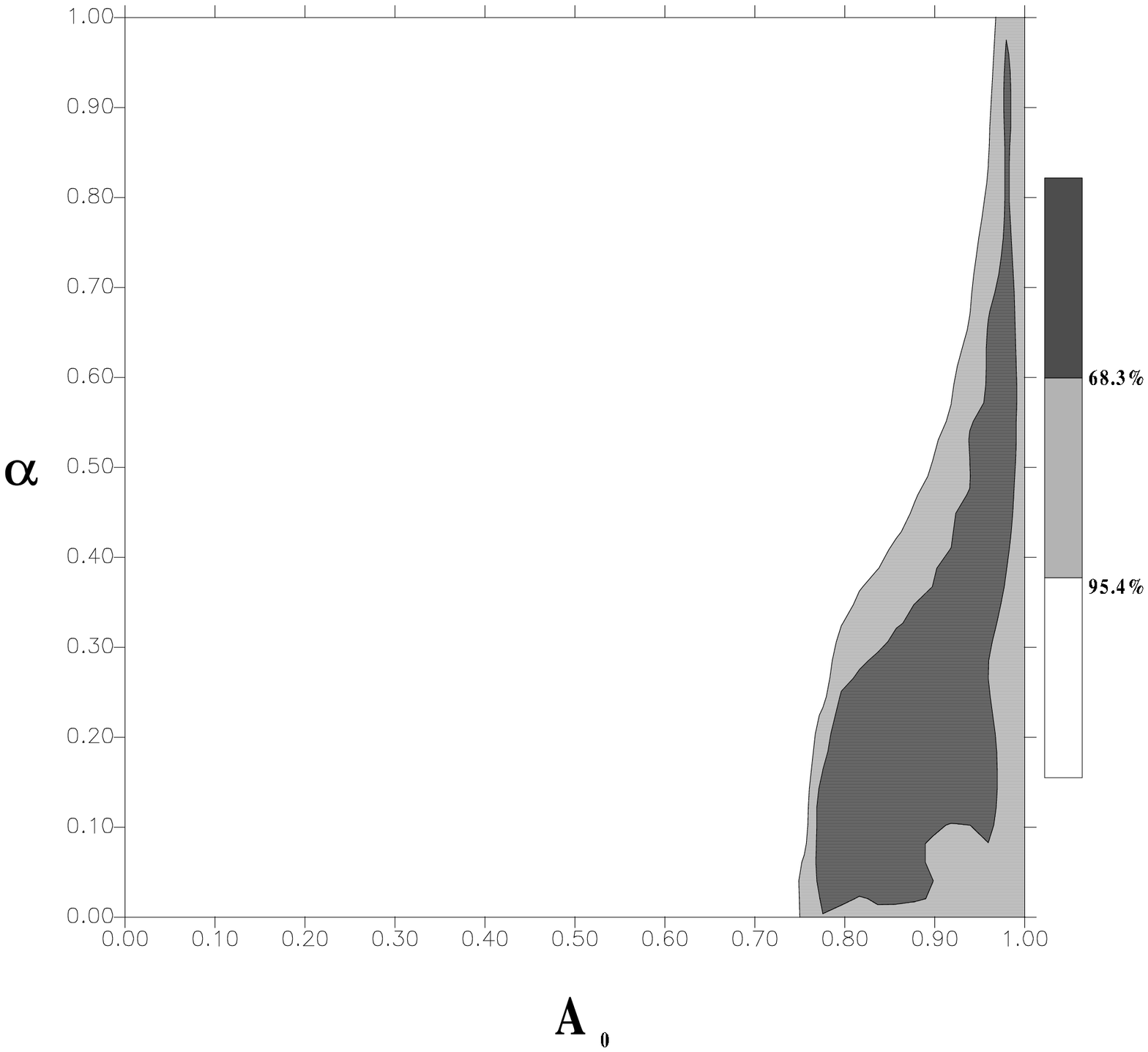}
\caption{Confidence  levels on the plane $(A_0,\alpha)$ for sample X1 ($\Lambda$CDM model) of
simulated SNAP data, marginalized over $\Omega_m$. The figure shows the ellipses of
preferred values of $A_0$ and $\alpha$.} \label{fig:26}
\end{figure}
 
\begin{figure}
\includegraphics[width=0.8\textwidth]{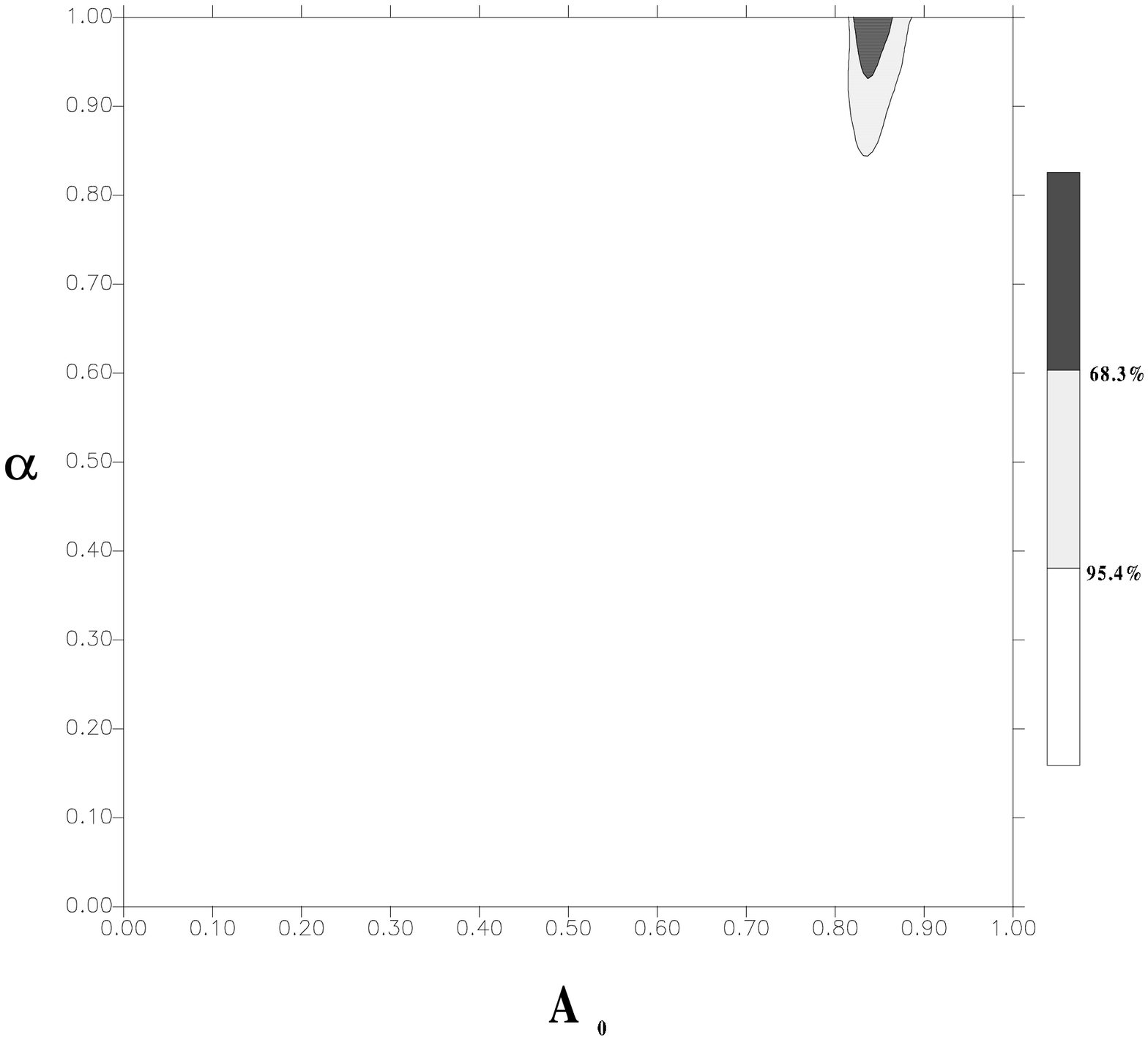}
\caption{Confidence levels on the plane $(A_0,\alpha)$ for sample X2
(Cardassian model) of simulated SNAP data, marginalized over $\Omega_m$. The figure shows the
ellipses of preferred values of $A_0$ and $\alpha$.} \label{fig:27}
\end{figure}

\begin{figure}
\includegraphics[width=0.8\textwidth]{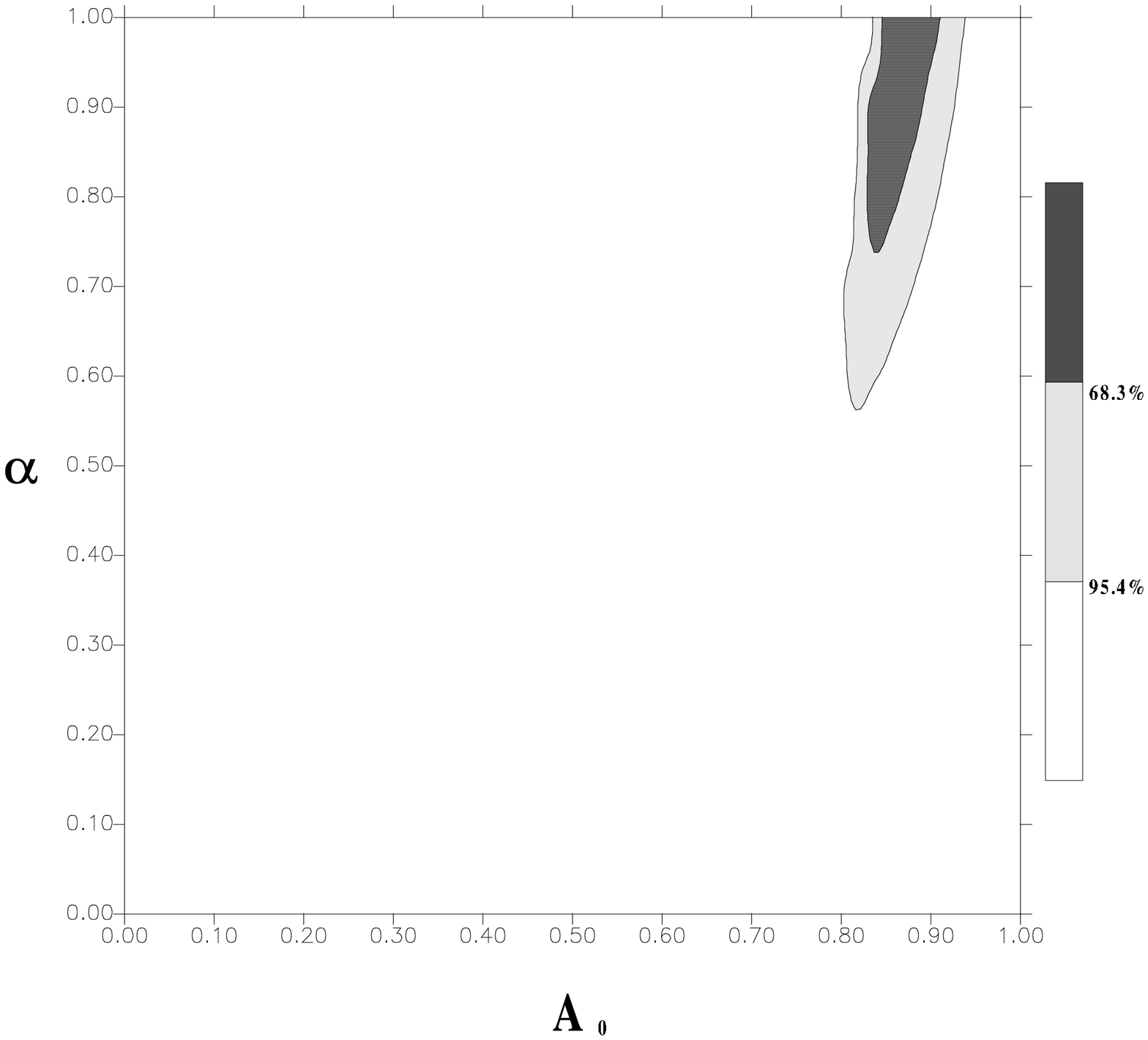}
\caption{Confidence  levels on the plane $(A_0,\alpha)$ for sample X3a (Generalized
Chaplygin Gas model) of simulated SNAP data ($\Omega_m=0$, $A_0=0.85$, $\alpha=1.$),
marginalized over $\Omega_m$. The figure shows the ellipses of preferred values of $A_0$
and $\alpha$.} \label{fig:28}
\end{figure}
 
\begin{figure}
\includegraphics[width=0.8\textwidth]{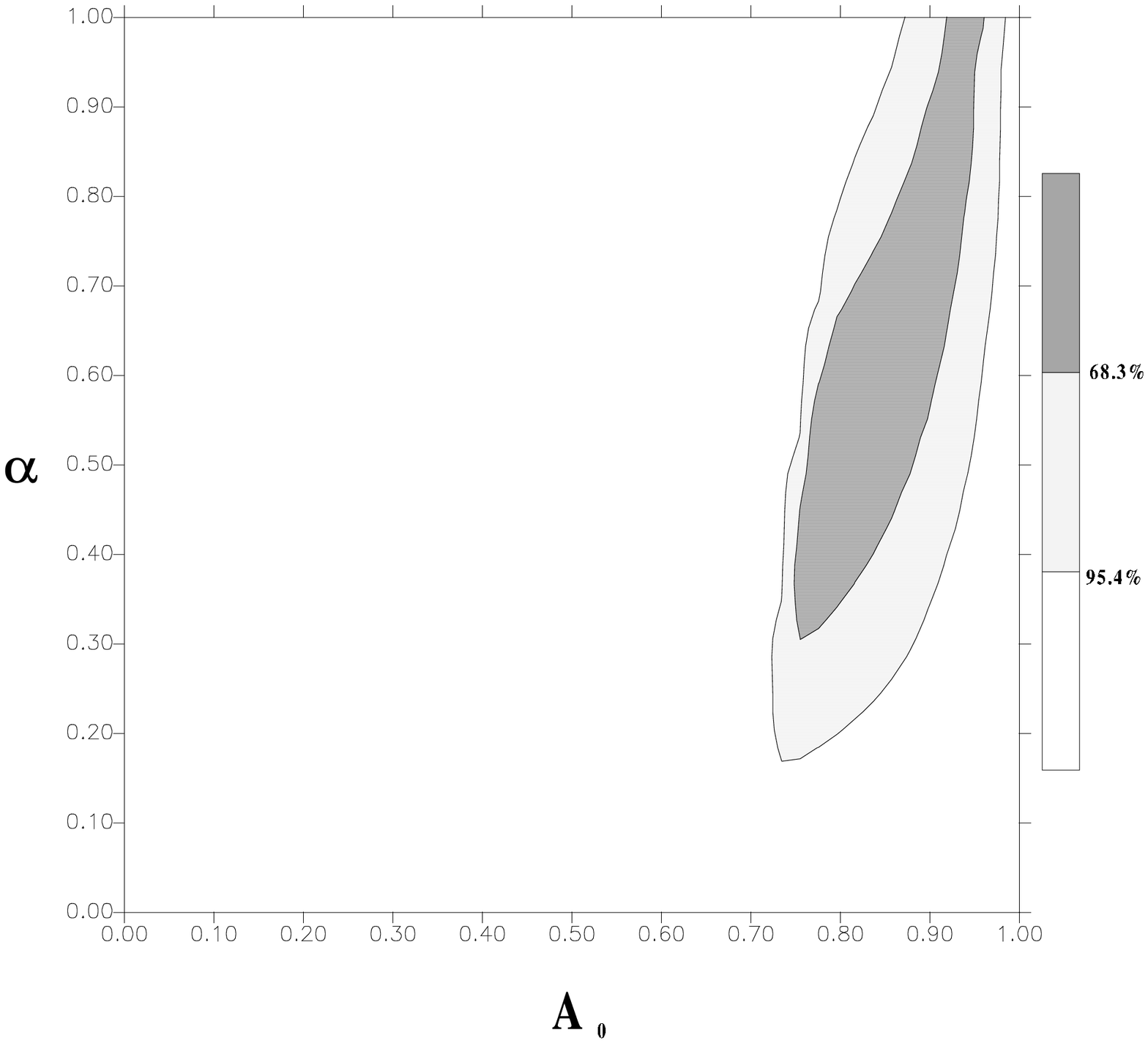}
\caption{Confidence  levels on the plane $(A_0,\alpha)$ for sample X3b
(Generalized Chaplygin Gas model) of simulated SNAP data
($\Omega_m=0$, $A_0=0.76$, $\alpha=0.49$), marginalized over $\Omega_m$.
The figure shows the ellipses of preferred values of $A_0$ and $\alpha$.}
\label{fig:29}
\end{figure}

\section{Acknowledgements}
We thank dr Barris and dr Riess for explanation of details of their SNIa samples. MS was
supported by KBN grant no 2P03D00326

\end{document}